\documentclass[aos,preprint]{imsart}
\RequirePackage[colorlinks,citecolor=blue,urlcolor=blue]{hyperref}

\usepackage{graphicx}

\usepackage{here}

\usepackage{latexsym}

\usepackage{amsbsy}

\usepackage{amsmath,amsthm,amssymb, amsfonts}

\newcommand{\notdse}{\not\hspace{-2mm}\dse}

\newcommand{\pa}{\mathrm{pa}}
\newcommand{\nei}{\mathrm{ne}}
\newcommand{\spo}{\mathrm{sp}}
\newcommand{\an}{\mathrm{an}}

\newcommand{\ant}{\mathrm{ant}}

\newcommand{\dse}{\,\mbox{$\perp$}\,}

\newcommand{\cip}{\mbox{\,$\perp\!\!\!\perp$\,}}

\newcommand{\cd}{\,|\,}

\newcommand{\nl}{\vspace{3mm}\\}

\newcommand{\nn}[0]{\hspace*{.7em}}

\newcommand{\node}{\mbox {\LARGE
{$\mbox{$\circ$}$}}}

\newcommand{\margn}{\mbox {\raisebox{-.1 ex}{\margnn}}}
\newcommand{\margnn}{\mbox {\Large
{$\not \: \not $}}$\node $}

\newcommand{\condnc}{\mbox{\raisebox{-.7ex}{\condnnc}}}
\newcommand{\condnnc}{\mbox {\LARGE{$\,
\mbox{$\Box$}
\raisebox{.21ex}{\hspace{-1.43ex}\mbox{$\circ$}}
$}} }

\newcommand{\ful}{\mbox{$\, \frac{ \nn \nn \;}{ \nn \nn
}$}}

\newcommand{\fla}{\mbox{$\hspace{.05em} \prec
\!\!\!\!\!\frac{\nn \nn}{\nn}$}}

\newcommand{\fra}{\mbox{$\hspace{.05em} \frac{\nn
\nn}{\nn
}\!\!\!\!\! \succ \! \hspace{.25ex}$}}

\newcommand{\arc}{\mbox{$\hspace{.06em} \prec
\!\!\!\!\!\frac{\nn \nn}{\nn}
\!\!\!\!\!
\succ\! \hspace{.25ex}$}}

\newtheorem{prop}{Proposition}
\newtheorem{coro}{Corollary}

\newtheorem{lemma}{Lemma}

\newtheorem{alg}{Algorithm}
\newtheorem{theorem}{Theorem}

\RequirePackage[OT1]{fontenc}
\RequirePackage{amsthm,amsmath,natbib}
\RequirePackage[colorlinks,citecolor=blue,urlcolor=blue]{hyperref}

\startlocaldefs
\endlocaldefs

\setcitestyle{numbers,square}

\begin{document}

\begin{frontmatter}
\title{Marginalization and Conditioning for LWF Chain Graphs}
\runtitle{Marginalization and Conditioning for LWF Chain Graphs}

\begin{aug}
\author{\fnms{Kayvan} \snm{Sadeghi}
\ead[label=e1]{k.sadeghi@statslab.cam.ac.uk}}

\address{Statistical Laboratory\\
Centre for Mathematical Studies\\
Wilberforce Road\\
Cambridge, CB3 0WB\\
United Kingdom.\\
  \printead{e1}}


\runauthor{K. Sadeghi}

\thankstext{t2}{Supported by grant $\#$FA9550-12-1-0392 from the U.S. Air Force Office of Scientific
Research (AFOSR) and the Defense Advanced Research Projects Agency (DARPA)}

\affiliation{University of Cambridge}

\end{aug}

\begin{abstract}
In this paper, we deal with the problem of marginalization over and conditioning on two disjoint subsets of the node set of chain graphs (CGs) with the LWF Markov property. For this purpose, we define the class of chain mixed graphs (CMGs) with three types of edges and, for this class, provide a separation criterion under which the class of CMGs is stable under marginalization and conditioning and contains the class of LWF CGs as its subclass. We provide a method for generating such graphs after marginalization and conditioning for a given CMG or a given LWF CG. We then define and study the class of anterial graphs, which is also stable under marginalization and conditioning and contains LWF CGs, but has a simpler structure than CMGs.
\end{abstract}

\begin{keyword}[class=AMS]
\kwd[Primary ]{62H99}
\kwd[; secondary ]{62A99}
\end{keyword}

\begin{keyword}
\kwd{$c$-separation criterion}
\kwd{chain graph}
\kwd{independence model}
\kwd{LWF Markov property}
\kwd{$m$-separation}
\kwd{marginalization and conditioning}
\kwd{mixed graph}
\end{keyword}

\end{frontmatter}

\section{Introduction} Graphical models use graphs, in which nodes are random variables and edges indicate some types of conditional dependencies. Mixed graphs, which are graphs with several types of edges, have started to play an important role in graphical models as they can deal with more complex independence structures that arise in different statistical studies.

The first example of mixed graphs in the literature appeared in \cite{lau89}. This was a chain graph (CG) with a specific interpretation of conditional independence, which is now generally known as the Lauritzen-Wermuth-Frydenberg or LWF interpretation.  A formal interpretation, i.e.\ a Markov property, was later provided by \cite{fry90}. This Markov property, together with other properties such as the factorization property was extensively discussed in \cite{lau96}. By the term LWF CGs, one refers to the class of CGs with a specific independence structure that comes from the LWF Markov property.

It has become apparent that CGs with the LWF interpretation of independencies are important tools in capturing conditional independence structure of various probability distributions. For example, \citet{stub98} showed that for every CG, there exists a strictly positive discrete probability distribution that embodies exactly the independence statements displayed by the graph, and \citet{pen11} proved that almost all the
regular Gaussian distributions that factorize with respect to a chain graph are faithful to it. This means that a Gaussian distribution chosen at
random to factorize as specified by the LWF CG will have the independence structure of the graph and will satisfy no more independence
constraints.

However, in the corresponding models to LWF CGs, when some variables are unobserved -- also called latent or hidden -- or when some variables are set to specific values, the implied independence structure, i.e.\ the corresponding independence structure after marginalization and conditioning respectively, is not well-understood.

The same problem for the well-known class of directed acyclic graphs (DAGs), which is a subclass of LWF CGs, has been a subject of study, and several classes of graphs have been defined in order to capture the marginal and conditional independence structure of DAGs. These include MC graphs \cite{kos02}, ancestral graphs \cite{ric02}, and summary graphs \cite{wer11}; see also \cite{sad13}. There is also a literature pertaining to this problem for other types of graphs; see, for example, the class of marginal AMP chain graphs in \cite{pen14} for marginalization in AMP chain graphs \cite{and01}.

For LWF CGs, as it will be shown in this paper, one can capture the independence structure induced by conditioning on some variables by another LWF CG, but in general cannot capture the independence structure induced by marginalization over some variables by a CG. In this sense, CGs are stable under conditioning but not under marginalization.

Indeed models with latent variables do not necessarily possess the desirable statistical properties of graphical models without latent variables, such as identifiability, existence of a unique MLE, or being curved exponential families in some cases such as DAGs; see, e.g.,\cite{gei01}.

However, a first step in dealing with this problem is, in the case of marginalization, to come up with a more complex class of graphs with a certain independence interpretation that captures the marginal independence structure of CGs; and in both cases of marginalization and conditioning, to provide methods by which the graphs that capture the marginal and conditional independence structure are generated. These are the main objectives of the current paper.

In the causal language (see, e.g., \cite{pea09}) the resulting classes of graphs give a simultaneous representation to ``direct effects", ``confounding", and ``non-causal symmetric dependence structures".

It is important to note that the classes of graphs introduced here only deals with the conditional independence constraints,
and not other constraints such as so-called \emph{Verma constraints} \cite{ver90}. The actual statistical model is much more complicated even when marginalizing DAGs; see, e.g., \cite{shp08}.

The introduction of these classes of graphs is also justified in the paper by showing that, for large subclasses of these classes of graphs, there are probability distributions (in fact both Gaussian and discrete) that are faithful to them. Although finding the explicit parametrizations for the definned graphs is beyond the scope of this paper, it also seems possible to extend the existing parametrizations for smaller types of graph in the literature to these classes in a fairly natural way. We will provide a discussion on this in the paper.

The structure of the paper is as follows: In the next section, we define mixed and chain graphs, and, for these classes of graphs, give graph theoretical definitions needed in this paper. In Section \ref{sec:3}, we provide two equivalent ways for reading off independencies from a CG based on the LWF Markov property.
In Section \ref{sec:5}, we define the class of chain mixed graphs with certain independence interpretation, and show that they capture the marginal independence structure of LWF CGs and that they are stable under marginalization, and provide an algorithm for generating such graphs after marginalization. In Section \ref{sec:6}, we show that the class of CMGs is also stable under conditioning, provide the corresponding algorithm, and combine marginalization and conditioning for CMGs. As a corollary, we see that LWF CGs are stable under conditioning. In Section \ref{sec:7}, we define the class of anterial graphs as a subclass of CMGs, which also contains LWF CGs, and show that this class is stable under marginalization and conditioning. We also provide an algorithm for marginalization and conditioning for this class. In Section \ref{sec:8}, we discuss the implications of the results for probabilistic independence models that are faithful to LWF CGs, and possible ways to generalize the parametrizations existing in the literature for CMGs and anterial graphs. In the Appendix in the supplementary material \cite{sad15s}, we provide proofs of non-trivial lemmas, propositions, and theorems in the paper as well as some more technical and yet less informative lemmas that are used in the proofs.
\section{Definitions for mixed graphs and chain graphs}
\subsection{Basic graph theoretical definitions} A \emph{graph} $G$ is a triple consisting of a \emph{node} set or
\emph{vertex} set $V$, an \emph{edge} set $E$, and a relation that with
each edge associates two nodes (not necessarily distinct), called
its \emph{endpoints}. When nodes $i$ and $j$ are the endpoints of an
edge, these are
\emph{adjacent} and we write $i\sim j$. We say the edge is \emph{between} its two
endpoints. We usually refer to a graph as an ordered
pair $G=(V,E)$. Graphs $G_1=(V_1,E_1)$ and $G_2=(V_2,E_2)$ are called \emph{equal} if $(V_1,E_1)=(V_2,E_2)$. In this case we write $G_1=G_2$.

Notice that graphs that we use in this paper (and in general in the context of graphical models) are so-called \emph{labeled graphs}, i.e.\ every node is considered a different object. Hence, for example, graph $i\ful j\ful k$ is not equal to $j\ful i\ful k$.

Here we introduce some
basic graph theoretical definitions. A \emph{loop} is an edge whose endpoints are equal. \emph{Multiple edges} are edges whose endpoints are the same as each other. A \emph{simple graph} has neither loops nor multiple edges. A \emph{complete} graph is a simple graph with all pairs of nodes adjacent.

A \emph{subgraph} of a graph $G_1$ is graph $G_2$ such that $V(G_2)\subseteq V(G_1)$ and $E(G_2)\subseteq E(G_1)$ and the assignment of endpoints to edges in $G_2$
is the same as in $G_1$. An \emph{induced subgraph} by a subset $A$ of the node set is a subgraph that contains the  node set $A$ and all edges between two
nodes in $A$.

A \emph{walk} is a list $\langle i_0,e_1,i_1,\dots,e_n,i_n\rangle$ of nodes and edges such that for $1\leq m\leq n$, the edge $e_m$ has endpoints $i_{m-1}$ and $i_m$. A \emph{path} is a walk with no repeated node or edge.  A \emph{cycle} is a walk with no repeated node or edge except $i_0=i_n$. If the graph is simple then a path or a cycle can be determined uniquely by an ordered sequence of nodes. Throughout this paper, however, we use node sequences to describe paths and cycles even in graphs with multiple edges, but we assume that the edges of the path are all determined. It is usually apparent from the context or the type of the path which edge belongs to the path in multiple edges.
We say a walk or a path is \emph{between} the first and the last nodes of the list in $G$. We
call the first and the last nodes \emph{endpoints} of the walk or of the path. All other nodes are the \emph{inner nodes}.

For a walk or path $\pi=\langle i_1,\dots.i_n\rangle$, any subsequence $\langle i_k,i_{k+1},\dots,i_{k+p}\rangle$, $1\leq k,k+p\leq n$, whose members appear consecutively on $\pi$, defines a \emph{subwalk} or a \emph{subpath} of $\pi$ respectively.

\subsection{Some definitions for mixed graphs}
A \emph{mixed graph} is a graph containing three types of edges denoted by
arrows, arcs (two-headed arrows), and lines (solid lines). Mixed
graphs may have multiple edges of different types but do
not have multiple edges of the same type.
We do not distinguish between $i\ful j$ and $j\ful i$ or $i\arc j$
and $j \arc i$, but we
do distinguish between $j\fra i$ and $i\fra j$. In this paper we are only considering mixed graphs that do not contain loops of any type. These constitute the class of \emph{loopless mixed graphs}.

For mixed graphs, we say that $i$ is a
\emph{neighbour} of $j$ if these are endpoints of a line, and $i$ is a parent of $j$ and $j$ is a child of $i$ if there is an arrow from $i$ to $j$. We also define that $i$ is a \emph{spouse} of $j$ if these are endpoints of an arc. We use the notations $\nei(j)$, $\pa(j)$, and $\spo(j)$ for the set of all neighbours, parents, and spouses of $j$ respectively.

In the cases of $i\fra j$ or
$i\arc j$ we say that there is an arrowhead pointing to (at) $j$.


A walk $\langle i=i_0,i_1,\dots,i_n=j\rangle$ is \emph{directed} from $i$ to $j$ if all $i_ki_{k+1}$ edges are arrows pointing from $i_k$ to $i_{k+1}$. If there is a directed walk from $j$ to $i$ then $j$ is an \emph{ancestor} of $i$ and $i$ is a \emph{descendant} of $j$. We denote the set of ancestors of $i$ by $\an(i)$. Notice that, unlike some authors,we do not consider $i$ to be in the set of ancestors or descendants of $i$. Moreover, a cycle with the above property is called a \emph{directed cycle}.

A walk $\langle i=i_0,i_1,\dots,i_n=j\rangle$ from $i$ to $j$ is a \emph{semi-directed walk} if it only consists of lines and arrows (it may contain only one type of edge), and every arrow $i_ki_{k+1}$ is pointing from $i_k$ to $i_{k+1}$. Thus a directed walk is a type of semi-directed walk. We shall say that $i$ is \emph{anterior} of $j$ if there is a semi-directed walk from $i$ to $j$. We use
the notation $\ant(i)$ for the set of all anteriors of $i$. Notice again that, similar to ancestors, we do not consider a node $i$ to be an anterior of itself. For a set of nodes $A$, we define $\ant(A)=\bigcup_{i\in A}\ant(i)\setminus A$. Notice also that, since ancestral graphs have no arrowheads pointing to lines, our definition of anterior extends the notion of anterior used in \cite{ric02} for ancestral graphs. Moreover, a cycle with the properties of semi-directed walks is called a \emph{semi-directed cycle}.

A \emph{section} of a walk in a mixed graph is a maximal subwalk that only consists of lines. Thus, any walk decomposes uniquely into sections (that are not necessarily edge-disjoint and may also be single nodes). Similar to nodes, all sections on a walk between $i$ and $j$ are \emph{inner sections} except those that contain $i$ or $j$, which are \emph{endpoint sections}.  As in any walk, we can also define the endpoints of a section. A section $\rho$ on a walk $\pi$ is called a  \emph{collider} section if one of the three following walks is a subwalk of $\pi$: $i\fra\rho\fla\,j$, $i\arc\rho\fla\,j$, and $i\arc\rho\arc\,j$. All other sections on $\pi$ are called \emph{non-collider} sections. We may speak of collider or non-collider sections without mentioning the relevant walk when this is apparent from context.

A \emph{trislide} on a walk $\pi$ is a subpath $\langle i=i_0,i_1,\dots,i_n=j\rangle$, where  $ii_1$ and $i_{n-1}j$ are arrows or arcs and the subpath $\rho'=\langle i_1,\dots,i_{n-1}\rangle$ is a section.

Three types of  trislides $i\fra\circ\ful\dots\ful\circ\fla\,j$, $i\arc\circ\ful\dots\ful\circ\fla\,j$, and $i\arc\circ\ful\dots\ful\circ\arc\,j$ are
\emph{collider} trislides  and all other types of trislides  are \emph{non-collider} on any walk of which the trislide is defined.

A \emph{tripath} is a trislide where the subpath $\rho'$ is a single node.  Note that \cite{sad13} used the term V-configuration for such a path. (\cite{kii84} and most texts let a V-configuration be a tripath with non-adjacent endpoints.) Tripaths and their inner nodes can be defined to be colliders or non-colliders as trislides and their inner sections.

Two walks $\pi_1$ and $\pi_2$ (including trislides, tripaths, or edges) between $i$ and $j$ are called \emph{endpoint-identical} if there is an arrowhead pointing to the endpoint section containing $i$ on $\pi_1$ if and only if there is an arrowhead pointing to the endpoint section containing $i$ on $\pi_2$; and similarly for $j$. For example, the paths $i\fra j$, $i\ful k\fra l\arc j$, and $i\fra k \arc l\ful j$ are all endpoint-identical as they have an arrowhead pointing to the section containing $j$ but  no arrowhead pointing to the section containing $i$ on the paths, but they are not endpoint-identical to $i\ful k\arc j$.
\subsection{Chain graphs}
\emph{Chain graphs} (CGs) is a graph consisting of lines and arrows that does not contain any semi-directed cycles with at least one arrow.

It is implied from the definition that CGs are
characterized by having a node set that can be partitioned into disjoint subsets
forming so-called \emph{chain components}. These are connected subgraphs consisting only of undirected edges and are obtained by removing all
arrows in the graph. All edges between nodes in the same chain component are lines, and all edges between different chain components are arrows. In addition, the chain components can be ordered in such a way that all arrows point from a chain with a higher number to one with a lower number.

\begin{figure}[H]
\centering
\begin{tabular}{cc}
\scalebox{0.15}{\includegraphics{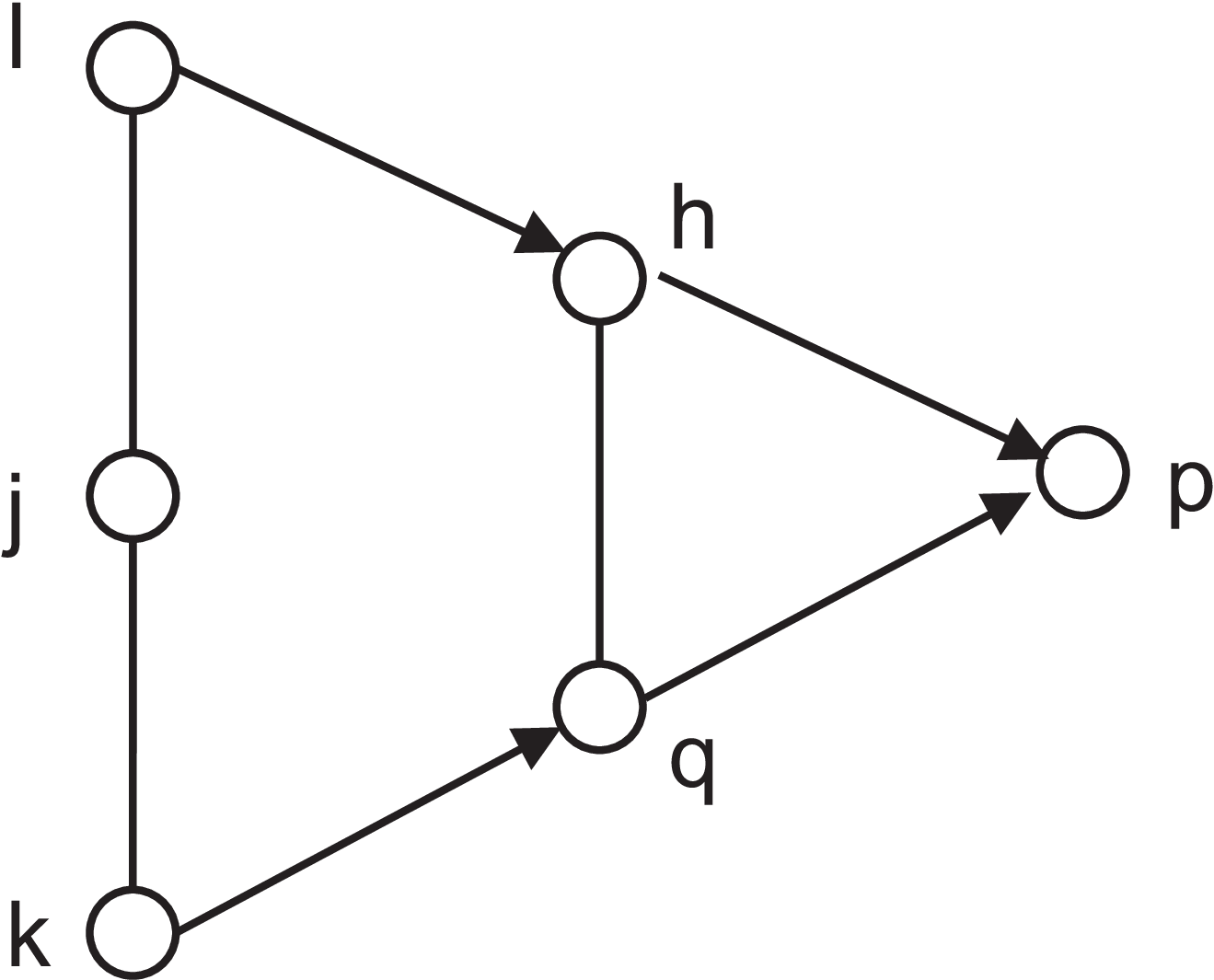}}\nn\nn &
\nn\nn\scalebox{0.15}{\includegraphics{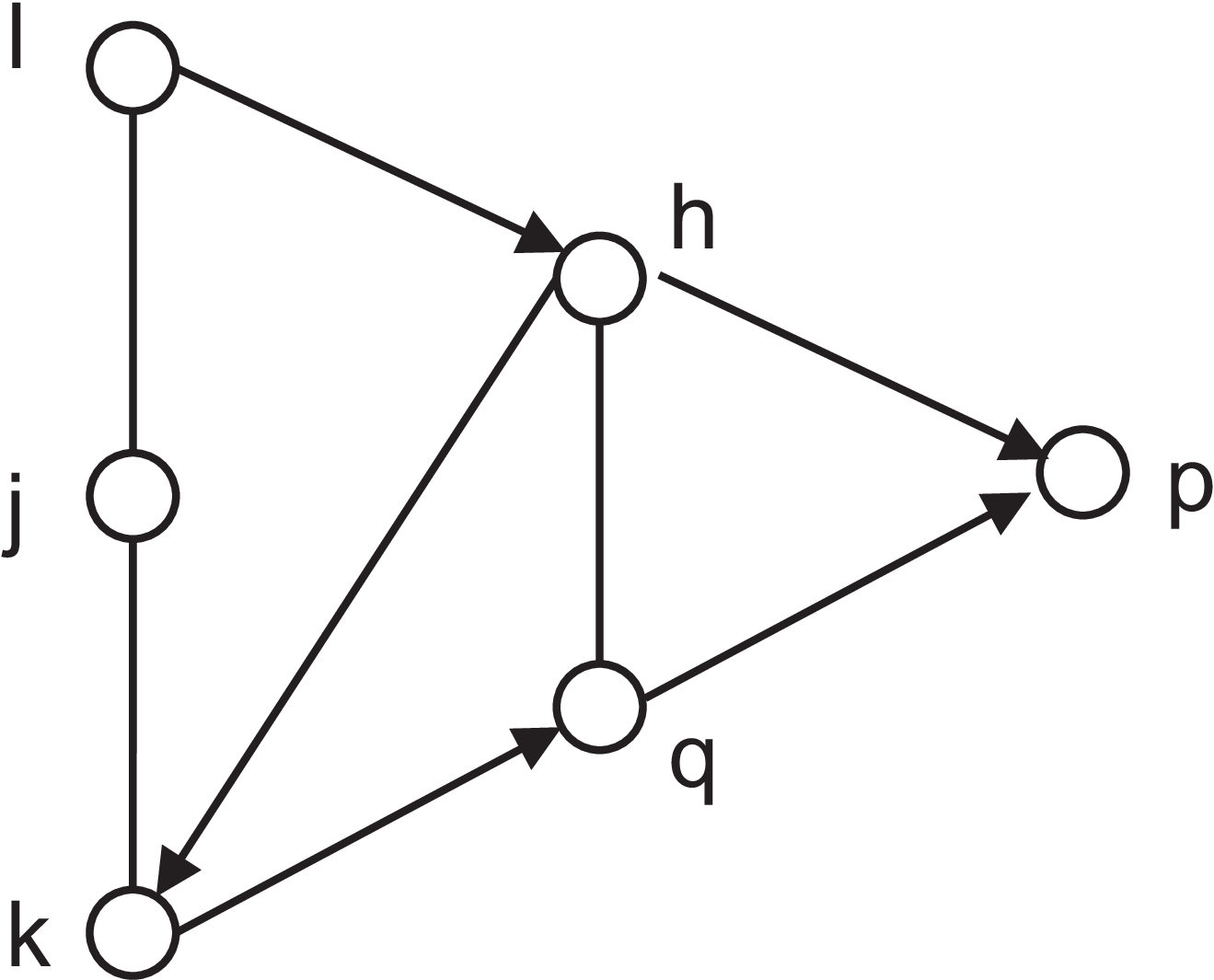}}\\
(a) & (b)
\end{tabular}
  \caption[]{\small{(a) A CG. (b) A mixed graph that is not a CG.}}
     \label{fig:chainex}
\end{figure}

For example, in Fig.\ \ref{fig:chainex}(a) the graph is a chain graph with chain components $\tau_1=\{l,j,k\}$, $\tau_2=\{h,q\}$, and $\tau_3=\{p\}$, but in Fig.\ \ref{fig:chainex}(b) the graph is not a chain graph because of the existence of the $\langle h,k,q\rangle$ semi-directed cycle.
If one replaces every chain component with a single node, one obtains a \emph{directed acyclic graph} (DAG), a graph consisting exclusively of arrows and without any directed cycles.

Notice that generally CGs are defined to contain arrows and one symmetric type of edge in their chain component, which can be. e.g., arcs. In this sense , the type of CG in which we are interested in this paper is a \emph{line CG}.
can be lines or arcs)
\section{LWF Markov property for CGs}\label{sec:3}
An \emph{independence model} $\mathcal{J}$ over a set $V$ is a set of triples $\langle X,Y\cd Z\rangle$ (called \emph{independence statements}), where $X$, $Y$, and $Z$ are disjoint subsets of $V$ and $Z$
can be empty, and $\langle \varnothing,Y\cd Z\rangle$ and $\langle X,\varnothing\cd Z\rangle$ are always included in $\mathcal{J}$. The independence statement $\langle X,Y\cd Z\rangle$ is interpreted as ``$X$ is independent of $Y$ given $Z$''. Notice that independence models contain probabilistic independence models as a special case. For further discussion on independence models, see \cite{stu05}.

A graph $G$ also induces an independence model $\mathcal{J}(G)$. One way is by using a \emph{separation criterion}, which determines whether for three disjoint subsets $A$, $B$, and $C$ of the node set of $G$, $\langle A,B\cd C\rangle\in \mathcal{J}(G)$. Such a criterion verifies whether $A$ is separated from $B$ by $C$ in the sense that  there are no walks or paths of specific types between $A$ and $B$ given $C$ in the graph. Such a separation is denoted by $A\dse B\cd C$. It is clear that $\mathcal{J}(G)$ satisfies the \emph{global Markov property}, which states that if $A\dse B\cd C$ in $G$ then  $\langle A,B\cd C\rangle\in \mathcal{J}$.

For CGs, at least four different separation criteria, i.e.\ four different types of global Markov property
have been discussed in the literature. \citet{drt09} has classified them as (1) the \emph{LWF} or \emph{block concentration} Markov property, (2) the \emph{AMP}
or \emph{concentration regression} Markov property, as defined and studied by \cite{and01}, (3) a Markov property that is dual to the AMP Markov property, and (4) the \emph{multivariate regression}
Markov property, as introduced by \cite{cox93} and studied extensively recently; for example see \cite{mar11,wers11}.

In this paper, we are interested in the LWF Markov property, and we introduce two equivalent separation criteria for this in this section. Henceforth, for the sake of brevity, by CGs we refer to CGs with the LWF Markov property.

The \emph{moralization criterion} for CGs was defined in \cite{fry90} and is a generalization of the moralization criterion for DAGs defined in \cite{lau88}; see also \cite{lau96}. The \emph{moral graph} of a chain graph $G$, denoted by $(G)^m$ is a graph that consists only of lines and that is generated from $G$ as follows: for every edge $ij$ in $G$ there is a line $ij$ in $(G)^m$. In addition if nodes $i$ and $j$ are parents of the same chain component in $G$ then there is the line $ij$ in  $(G)^m$.

Now let $G_{\ant(A\cup B\cup C)}$ be the induced subgraph of $G$ generated by $\ant(A\cup B\cup C)$. The moralization criterion states that for $A$, $B$, and $C$, three disjoint subsets of the node set of $G$, if there are no paths between $A$ and $B$ in $(G_{\ant(A\cup B\cup C)})^m$ whose inner nodes are outside $C$ then $A\dse_{mor} B\cd C$.

An equivalent criterion, called the \emph{$c$-separation criterion} for CGs was defined in \cite{stub98}. Here we present a simpler version of that criterion, presented in \cite{stu98}, with a different notation and wording:

A walk $\pi$ in a CG is a $c$-connecting walk given $C$ if every collider section of $\pi$ has a node in $C$ and all  non-collider sections are outside $C$.
A section on $\pi$ is \emph{open} if either: it is a collider section and one of its nodes is in $C$; or
it is a non-collider section and all its nodes are outside $C$. Otherwise it is \emph{blocked}. We say that $A$ and $B$ are $c$-separated given $C$ if there are no $c$-connecting walks between $A$ and $B$ given $C$, and we use the notation $A\dse_c B\cd C$.

Notice that, as mentioned in \cite{stub98}, there is potentially an infinite number of walks, and therefore, this might not be an appropriate criterion for testing independencies. Although, in this paper, we only use this criterion in order to prove our theoretical results regarding marginalization and conditioning, and an infinite number of walks is not an issue for this purpose, in \cite{stu98}, it was shown that this criterion can also be implemented with an algorithm.

\begin{figure}
\centering
\begin{tabular}{cc}
\scalebox{0.22}{\includegraphics{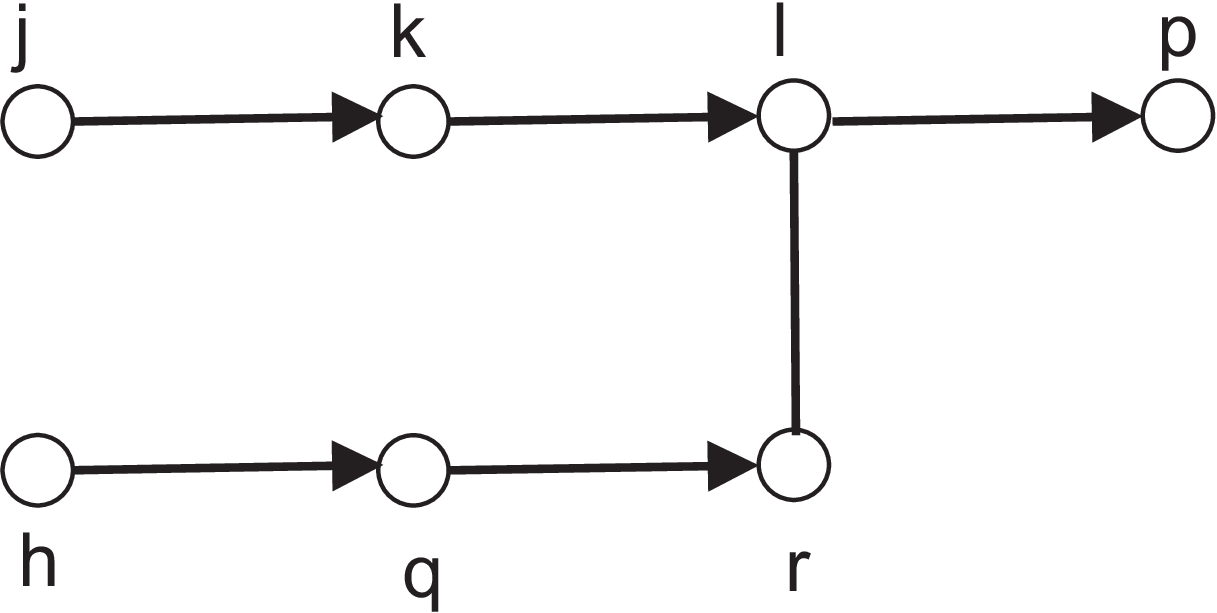}}\nn\nn &
\nn\nn\scalebox{0.22}{\includegraphics{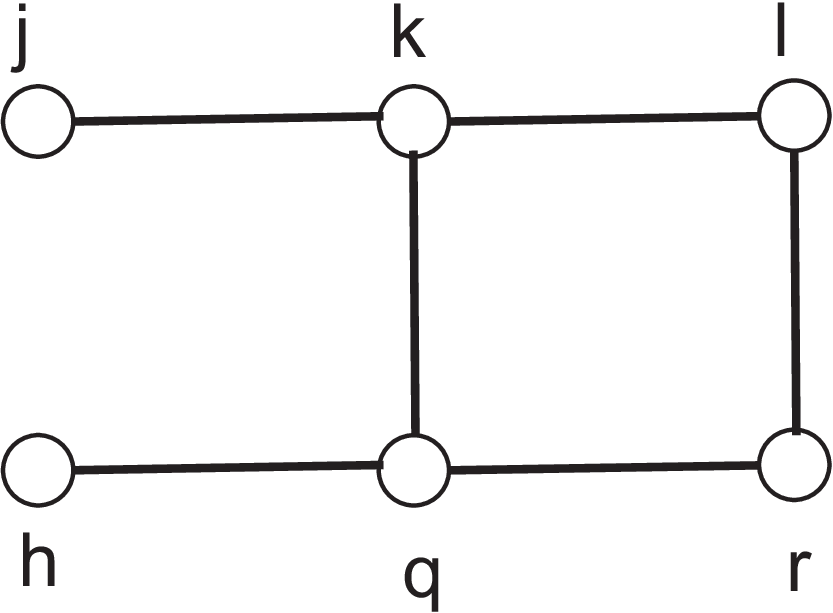}}\\
(a) & (b)
\end{tabular}
  \caption[]{\small{(a) A chain graph $G$. (b) The moral graph $(G_{\ant(\{j,h,l\})})^m$.}}
     \label{fig:LWFex}
\end{figure}

For example, in the graph of Fig.\ \ref{fig:LWFex}(a), the independence statement $j\dse h\cd l$ does not hold. This can be seen by looking at the moral graph  $(G_{\ant(\{j,h,l\})})^m=(G_{\{j,h,k,q,l,r\}})^m$ in Fig.\ \ref{fig:LWFex}(b), and observing that the inner nodes of the path $\langle j,k,q,h\rangle$ are outside the conditioning set. The same conclusion can be made by looking at the walk $\langle j,k,l,r,q,h\rangle$, where the non-collider sections $k$ and $q$ are outside the conditioning set, but the inner node $l$ of the collider section $\langle l,r\rangle$ is in the conditioning set.

The equivalence of the moralization criterion and the original $c$-separation criterion was proven in Consequence 4.1 in \cite{stub98}. The equivalence with the mentioned simplified criterion was proven in \cite{stu98}. We use the notation $\mathcal{J}_c(G)$ for the independence model induced from $G$ by the above criteria.

We first prove the following lemma, which provides an equivalent type of walk to $c$-connecting walks:
\begin{lemma}\label{lem:0}
There is a $c$-connecting walk between $i$ and $j$ given $C$ if and only if there is a walk between $i$ and $j$ whose sections are all paths, and on which nodes of every collider section are in $C\cup\ant(C)$, and non-collider sections are outside $C$. In addition, these walks can be chosen to be endpoint-identical.
\end{lemma}
Notice that by the same method as the proof of this lemma, one can always assume that a section on a walk is a path. This is our assumption throughout the paper unless otherwise stated.
\section{Stability of CGs under marginalization and conditioning}\label{sec:5}
 For a subset $C$ of $V$, the \emph{independence model after conditioning on $C$}, denoted by $\alpha(\mathcal{J};\varnothing,C$), is
\begin{displaymath}
\alpha(\mathcal{J};\varnothing,C)=\{\langle A,B\cd D\rangle:\langle A,B\cd D\cup C\rangle\in \mathcal{J}\text{ and }(A\cup B\cup D)\cap C=\varnothing\}.
\end{displaymath}

One can observe that $\alpha(\mathcal{J};\varnothing,C)$ is an independence model over $V\setminus C$.

We now present the definition of \emph{stability under conditioning} \cite{sad13}: Consider a family of graphs $\mathcal{T}$. If, for every graph $G=(V,E)\in\mathcal{T}$ and every disjoint subsets $C$ of $V$, there is a graph $H\in\mathcal{T}$ such that $\mathcal{J}(H)=\alpha(\mathcal{J}(G);\varnothing,C)$ then $\mathcal{T}$ is stable under conditioning. Notice that the node set of $H$ is $V\setminus  C$.

We will see as a corollary of the results and algorithms in the next section that CGs are stable under conditioning.

Similar to the conditioning case, for a subset $M$ of $V$, the \emph{independence model after marginalization over $M$}, denoted by $\alpha(\mathcal{J};M,\varnothing$), is defined by
\begin{displaymath}
\alpha(\mathcal{J};M,\varnothing)=\{\langle A,B\cd D\rangle\in \mathcal{J}:(A\cup B\cup D)\cap M=\varnothing\}.
\end{displaymath}

One can observe that $\alpha(\mathcal{J};M,\varnothing)$ is an independence model over $V\setminus M$.

The definition of \emph{stability under marginalization} is defined similarly to the conditioning case: for a family of graphs $\mathcal{T}$, if, for every graph $G=(V,E)\in\mathcal{T}$ and every disjoint subsets $C$ of $V$, there is a graph $H\in\mathcal{T}$ such that $\mathcal{J}(H)=\alpha(\mathcal{J}(G);M,\varnothing)$ then $\mathcal{T}$ is stable under marginalization. We see again that the node set of $H$ is $N=V\setminus  M$.

\begin{figure}[H]
\centering
\scalebox{0.51}{\includegraphics{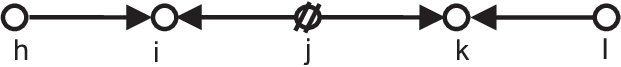}}
  \caption[]{\small{(a) A chain graph $G$, by which it can be shown that the class of CGs is not stable under marginalization. ($\margn\in M$.)}}
     \label{fig:margcountex}
\end{figure}

CGs are not closed under marginalization. For example, it can be shown that $G$ in Fig.\ \ref{fig:margcountex} is a CG (in fact a DAG) whose induced marginal independence model cannot be represented by a CG. We leave the details as an exercise to the reader.

Hence, we define a class of graphs that is stable under marginalization and contains CGs: the class of \emph{chain mixed graphs} (CMGs) is the class of mixed graphs without semi-directed cycles with at least an arrow. Notice that we allow CMGs to have multiple edges consisting of arcs and arrows and arcs and lines. This is a generalization of chain graphs since if a CMG does not contain arcs then it is a chain graph.

\begin{figure}[H]
\centering
\begin{tabular}{cc}
\scalebox{0.15}{\includegraphics{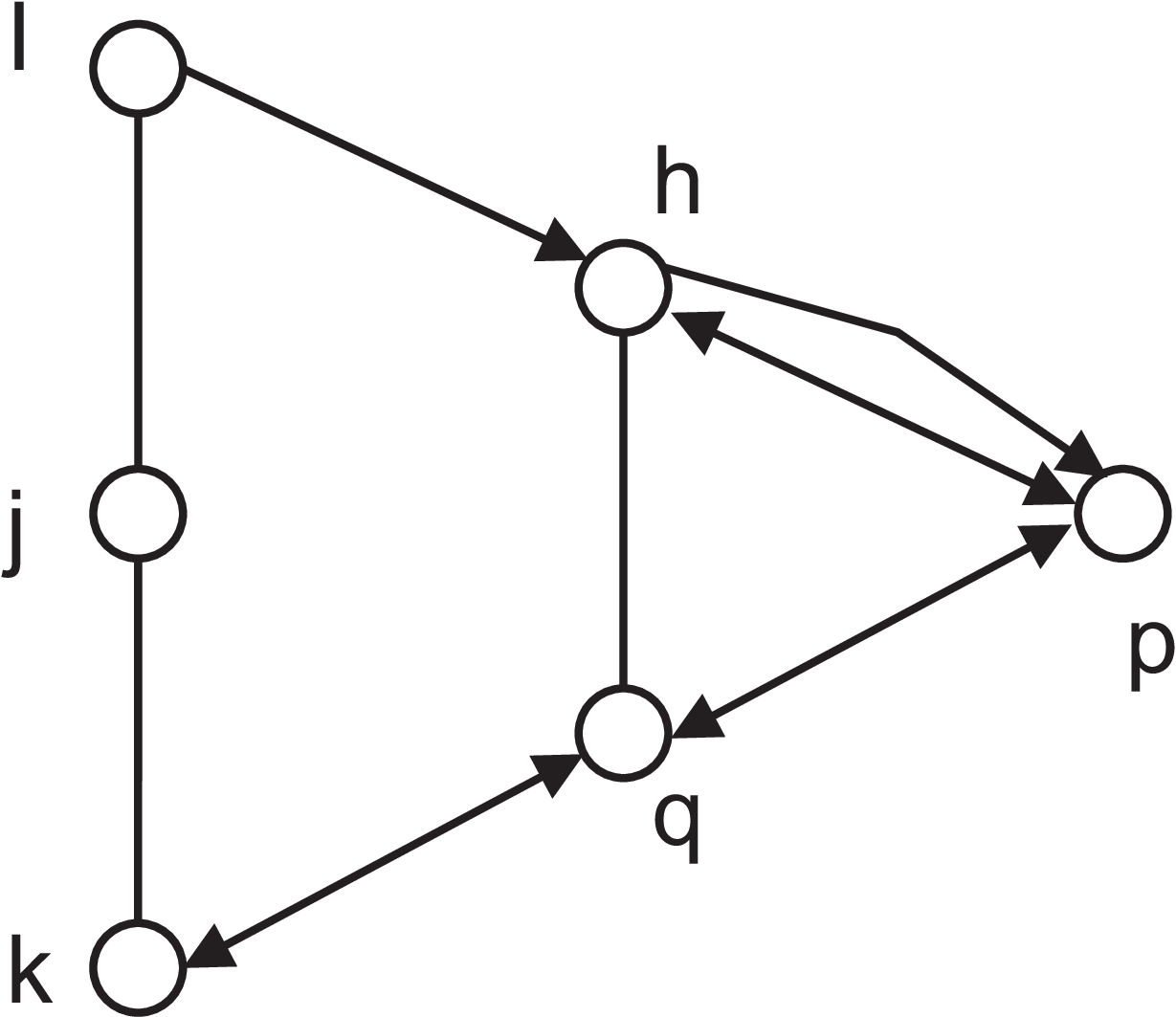}}\nn\nn &
\nn\nn\scalebox{0.15}{\includegraphics{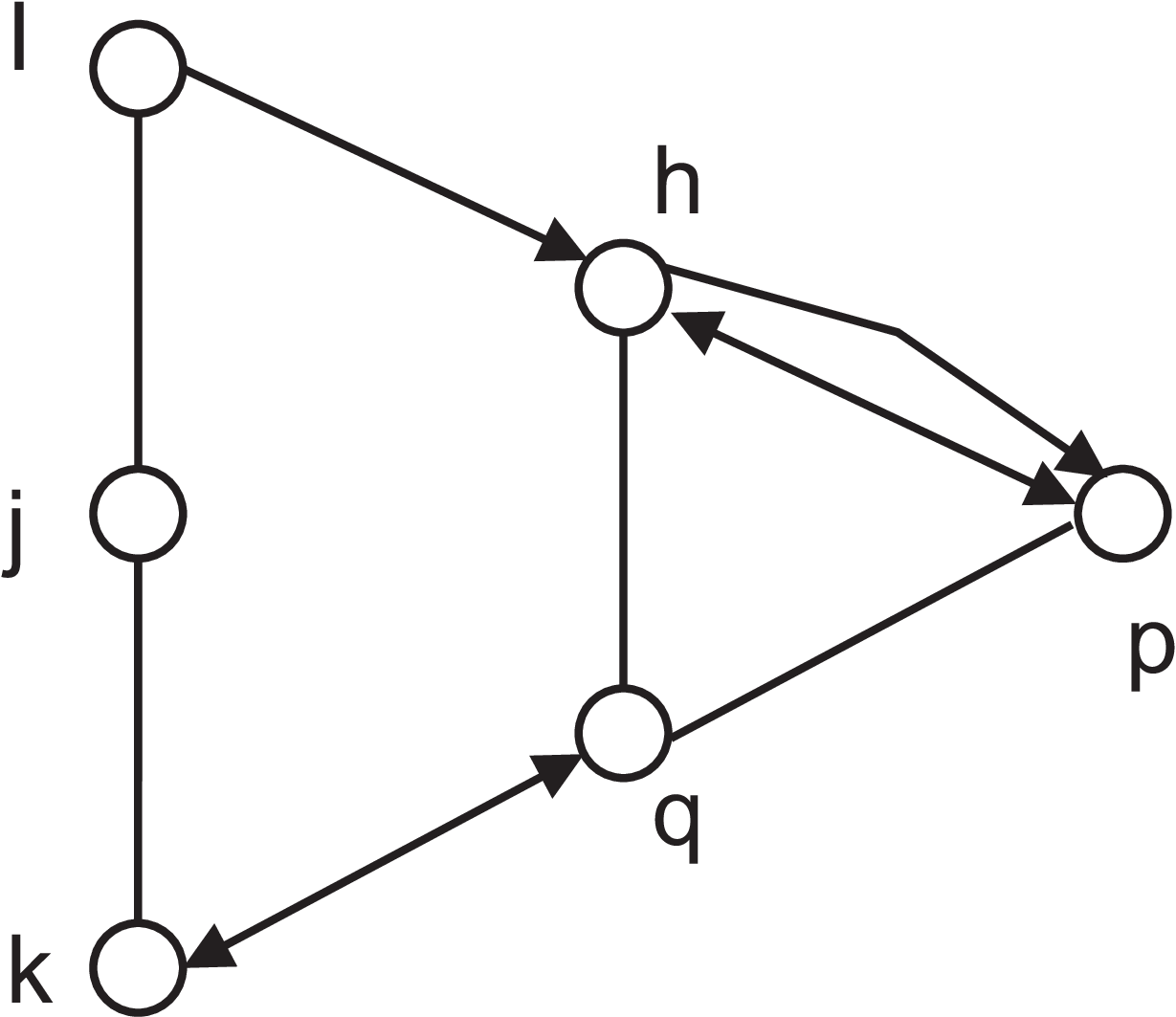}}\\
(a) & (b)
\end{tabular}
  \caption[]{\small{(a) A CMG. (b) A mixed graph that is not a CMG.}}
     \label{fig:CMGex}
\end{figure}

For example, in Fig.\ \ref{fig:CMGex}(a) the graph is a CMG, but in Fig.\ \ref{fig:CMGex}(b) the graph is not a CMG because of the existence of the $\langle h,p,q\rangle$ semi-directed cycle.


We provide a $c$-separation criterion for CMGs, and using this, show that CMGs are closed under marginalization. For this purpose, we provide in this section an algorithm that, from a CMG (or a chain graph) $G$ and after marginalization over $M$, generates a CMG with the corresponding independence model after marginalization over $M$.

We define a $c$-separation criterion for CMGs with exactly the same wordings as that of CGs: a walk $\pi$ in a CG is a $c$-connecting walk given $C$ if every collider section of $\pi$ has a node in $C$ and all  non-collider sections are outside $C$. We say that $A$ and $B$ are $c$-separated given $C$ if there are no $c$-connecting walks between $A$ and $B$ given $C$, and we use the notation $A\dse_c B\cd C$.

However, notice that this is in fact a generalization of the $c$-separation criterion for CGs since, for CMGs,  bidirected edges on $\pi$ may make a section collider.

We now provide an algorithm that, from a chain mixed graph $G$ and after marginalization over $M$, generates a CMG with the corresponding independence model after marginalization over $M$. Notice that this algorithm may indeed be applied to a CG.

 \begin{alg}\label{alg:2}
$\alpha_{CMG}(G;M,\varnothing)$:(\text{Generating a CMG
from a chain}\\
\text{mixed graph $G$ after marginalization over $M$)}\\
Start from $G$.
\begin{enumerate}
    \item Generate an $ij$ edge as in  Table \ref{tab:1}, steps 8 and 9, between $i$ and $j$ on a collider trislide with an endpoint $j$ and an endpoint in $M$ if the edge of the same type does not already exist.
    \item Generate an appropriate edge as in Table \ref{tab:1}, steps 1 to 7, between the endpoints of every tripath
    with inner node in $M$ if the edge of the same type does not already exist. Apply this step until no other edge can be generated.
    \item  Remove all nodes in $M$.
\end{enumerate}
\end{alg}
\begin{table}[H]
\caption{\small{Types of edge induced by tripaths with inner node $m\in M$ and trislides with endpoint $m\in M$.}}\label{tab:1}
\vspace{5mm}
\centering
\begin{tabular}{|c|c|c|c|}
    \hline
    1 & $i\fla m\fla\, j$ & generates & $i\fla j$\nl
    2 & $i\fla m\ful\, j$ & generates & $i\fla j$\nl
    3 & $i\arc m\ful\, j$ & generates & $i\arc j$\nl
    4 & $i\fla m\fra\, j$ & generates & $i\arc j$\nl
    5 & $i\fla m\arc\, j$ & generates & $i\arc j$\nl
    6 & $i\ful m\fla\, j$ & generates & $i\fla j$\nl
    7 & $i\ful m\ful\, j$ & generates & $i\ful j$\nl
    \hline
    8 & $m\fra i\ful\cdots\ful \circ\fla\, j$ & generates & $i\fla j$ \nl
    9 & $m\fra i\ful\cdots\ful \circ\arc\, j$ & generates & $i\arc j$ \\
    \hline
\end{tabular}
\\
\end{table}
Notice that, here and elsewhere, by removing nodes we mean also removing all the adjacent edges to those nodes. Notice also that all the cases generate an endpoint-identical edge to the tripath or the trislide. In addition, in cases 8 and 9, the node $m$ is separate  from the inner nodes of the concerned trislide since otherwise there will be a semi-directed cycle in the graph.

This algorithm is a generalization of the marginalization part of the summery-graph-generating algorithm \cite{sad13}. The first seven cases are exactly the same as the corresponding cases in the summery-graph-generating algorithm, whereas cases $8$ and $9$ do not appear in the summery-graph-generating algorithm since in summary graphs there are no arrowheads pointing to lines. The other reason is that here we deal with connecting walks instead of paths, and the subwalk $\langle i,m,i\rangle$ may be present in a connecting walk. In general, here in this algorithm, and in later algorithms in this paper, the sections are treated in the same way as the nodes are treated in the algorithms that generate summary graphs, acyclic directed mixed graphs (ADMGs) \cite{ric03}, or ancestral graphs. It is also worth noticing that all these algorithms are indeed generalizations of the ordinary latent projection operation; see \cite{pea09}.

\begin{figure}
\centering
\begin{tabular}{cc}
\scalebox{0.22}{\includegraphics{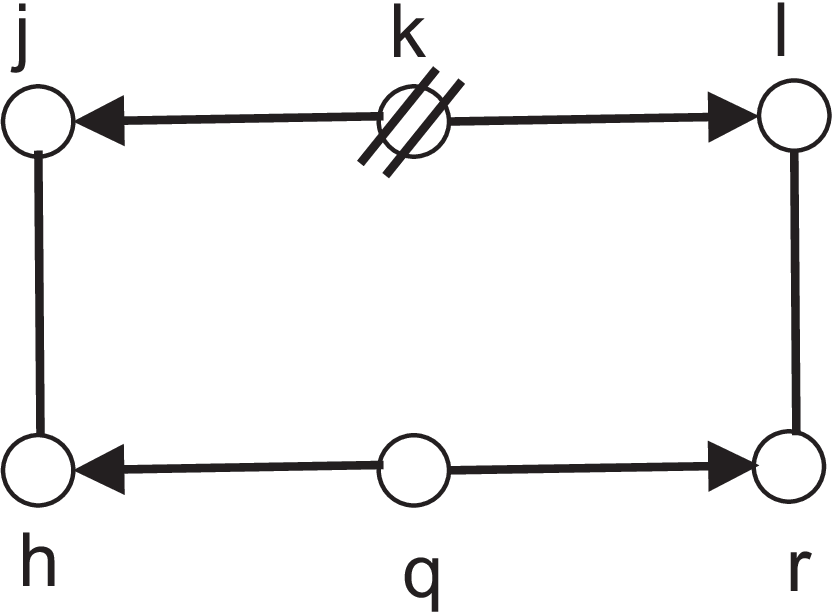}} &
\scalebox{0.22}{\includegraphics{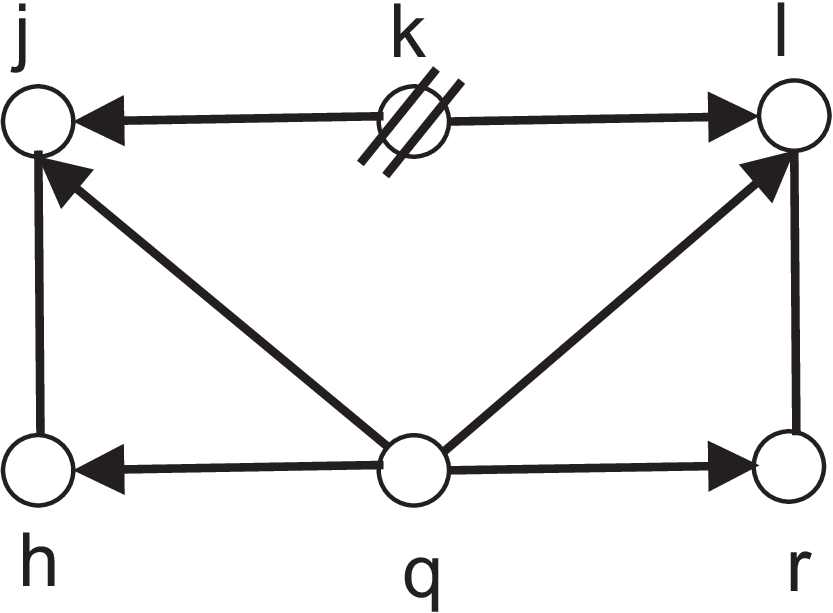}}\\
(a) & (b)\\
\scalebox{0.22}{\includegraphics{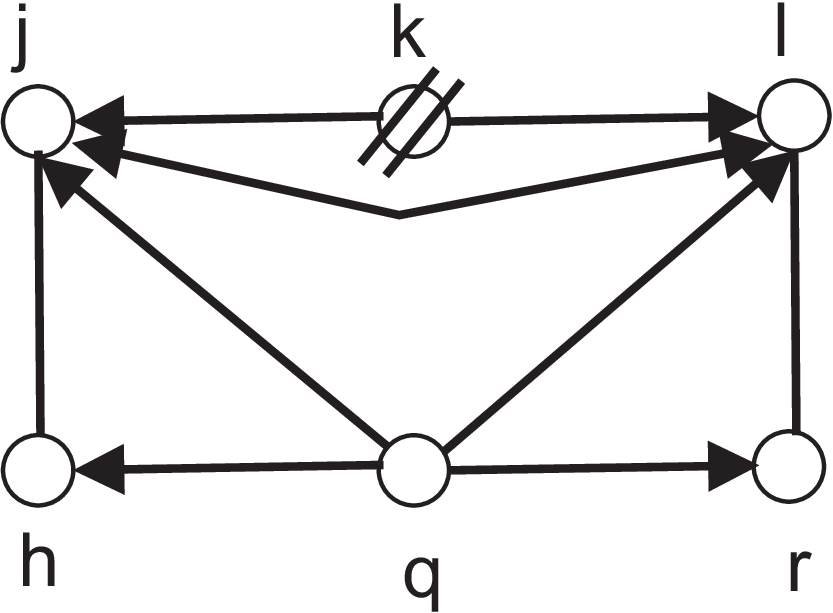}} &
\scalebox{0.22}{\includegraphics{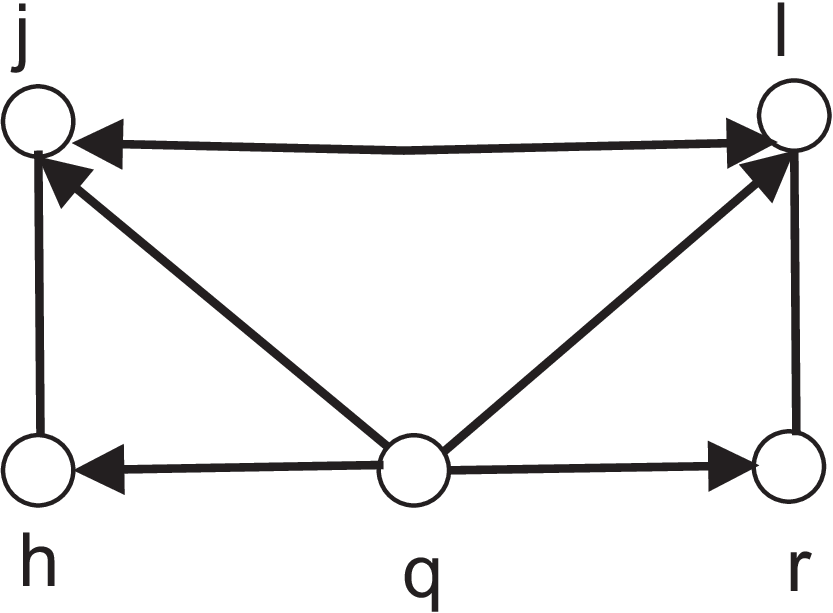}}
\\
(c) & (d)
\end{tabular}
  \caption[]{\small{(a) A chain graph $G$, $\margn \in M$. (b) The graph after applying step 1 of Algorithm \ref{alg:2} (case 8 of Table \ref{tab:1}). (c) The graph after applying step 2 of Algorithm \ref{alg:2} (case 4 of Table \ref{tab:1}) .
  (d) The generated CMG after applying step 3.}}
     \label{fig:alg2ex}
\end{figure}

Fig.\ \ref{fig:alg2ex} illustrates how to apply Algorithm \ref{alg:2} step by step to a CG.
We consider Algorithm \ref{alg:2} a function denoted by $\alpha_{CMG}$. Notice that for every chain mixed graph $G$, it holds that $\alpha_{CMG}(G;\varnothing,\varnothing)=G$. We first show that $\alpha_{CMG}(G;M,\varnothing)$ is a CMG:
\begin{prop}\label{prop:2}
Graphs generated by Algorithm \ref{alg:2} are CMGs.
\end{prop}
We first provide lemmas that express the global behavior of step 2 of Algorithm \ref{alg:2} as well as a generalization and an implication of step 1 (in the Appendix in \cite{sad15s}):
\begin{lemma}\label{lem:1}
Let $G$ be a CMG. There exists an edge between $i$ and $j$ in $\alpha_{CMG}(G;M,\varnothing)$ if and only if there exists an endpoint-identical walk between $i$ and $j$ in the graph generated after applying step 1 of Algorithm \ref{alg:2} to $G$ whose inner sections are all non-collider and whose inner nodes are all in $M$.
\end{lemma}
The following theorem shows that $\alpha_{CMG}(\cdot; \cdot,\varnothing)$ is well-defined in the sense that, instead of directly generating a CMG, we can split the nodes that
we marginalize over into two parts, first generate the CMG related to the first
 part, then from the generated CMG, generate the desired CMG related to the second part.
\begin{theorem}\label{thm:2n}
For a chain mixed graph $G$ and disjoint subsets $M$ and $M_1$ of its node set,
\begin{displaymath}
\alpha_{CMG}(\alpha_{CMG}(G;M,\varnothing);M_1,\varnothing)=\alpha_{CMG}(G;M\cup M_1,\varnothing).
\end{displaymath}
\end{theorem}
Some CMGs may not be generated after marginalization for CGs.  In the following proposition, we provide the exact set of graphs to which CMGs are mapped after marginalization. Denote by $\mathcal{CG}$ the set of all CGs and by $\mathcal{CMG}$ the set of all CMGs.
\begin{prop}\label{prop:2vn}
Define $\mathcal{H}$ to be the subset of $\mathcal{CMG}$ with the following properties:
\begin{enumerate}
  \item There is no collider trislide of form $k\arc i\ful\dots\ful j\fla l$ unless there is an arrow from $l$ to $i$;
  \item there is no collider trislide of form $k\arc i\ful\dots\ful j\arc l$ unless there are $kj$, $il$, and $ij$ arcs.
\end{enumerate}
Then $\alpha_{CMG}(\cdot;\cdot,\varnothing)$ maps $\mathcal{CG}$ and a subset of the node set of its member surjectively onto $\mathcal{H}$.
\end{prop}
Here we prove the main result of this section:
\begin{theorem}\label{thm:2}
For a chain mixed graph $G$ and disjoint subsets $A$, $B$, $M$, and $C_1$ of its node set,
\begin{displaymath}
\langle A,B\cd C_1\rangle\in\mathcal{J}_c(\alpha_{CMG}(G;M,\varnothing)) \iff \langle A,B\cd C_1\rangle\in\mathcal{J}_c(G).
\end{displaymath}
\end{theorem}
We, therefore, have the following immediate corollary:
\begin{coro}
The class of chain mixed graphs, $\mathcal{CMG}$, with $c$-separation criterion is stable under marginalization.
\end{coro}\label{coro:2}
\section{Stability of CMGs under marginalization and conditioning}\label{sec:6}
\subsection{Stability of CMGs under conditioning} In the previous section, we showed that the class of CMGs is stable under marginalization. In this section, we first show that the class of CMGs is also stable under conditioning, and provide an algorithm for conditioning for CMGs:
\begin{alg}\label{alg:3}
$\alpha_{CMG}(G;\varnothing,C)$:(\text{Generating a CMG
from a chain mixed}\\
\text{graph $G$ after conditioning on $C$)}\\
Start from $G$.
\begin{enumerate}
    \item Find all nodes in $C\cup\ant(C)$ and call this set $S$.
    \item  For collider trislides illustrated in Table \ref{tab:2}, steps 4 and 5, with an endpoint $i$ and one endpoint in $S$, generate an $ij$ edge following the table if the edge does not already exist.
    \item  For collider trislides (including tripaths) illustrated in Table \ref{tab:2}, steps 1-3, with at least one inner node in $S$, generate an edge following the table if the edge does not already exist. Apply this step repeatedly until no other edge can be generated, but do not use generated lines (to generate new sections).
    \item Remove the arrowheads of all arrows and arcs pointing to members of $S$ (i.e.\ turn such arrows into lines and such arcs into arrows).
\item Remove all nodes in $C$.
\end{enumerate}
\end{alg}
\begin{table}[H]
\caption{\small{Types of edges induced by trislides with an inner node or endpoint $s\in S=C\cup \ant(C)$.}}\label{tab:2}
\vspace{5mm}
\centering
\begin{tabular}{|c|c|c|c|}
    \hline
    1 & $i\fra s\ful\cdots\ful s\fla\, j$ & generates & $i\ful j$ \nl
    2 & $i\arc s\ful\cdots\ful s\fla\, j$ & generates & $i\fla j$ \nl
    3 & $i\arc s\ful\cdots\ful s\arc\, j$ & generates & $i\arc j$\nl
    \hline
    4 & $s\arc i\ful\cdots\ful \circ\fla\, j$ & generates & $i\fla j$ \nl
    5 & $s\arc i\ful\cdots\ful \circ\arc\, j$ & generates & $i\arc j$ \\
    \hline
\end{tabular}
\\
\end{table}
Notice that if a node of a section is in $S$ then all the inner nodes are in $S$, thus, we may speak of a section being in $S$. Notice also that all the steps of the algorithm generate endpoint-identical edges to the concerned trislides. In addition, we can assume that the endpoints of trislides are disjoint from the inner nodes, since (1) $j$ as an endpoint of an arrow cannot be also an inner node because the graph does not contain semi-directed cycles; and (2)  cases 2 and 3 with $i$ an inner node are equivalent to cases 4 and 5 respectively, and  cases 4 and 5 with $s$ an inner node are equivalent to cases 2 and 3 respectively.

Similar to Algorithm \ref{alg:2}, this algorithm is a generalization of the conditioning part of the summery-graph-generating algorithm \cite{sad13}. The first three cases are the same when one considers sections here to be the nodes in the summery-graph-generating algorithm. Cases $4$ and $5$ do not appear in the summery-graph-generating algorithm for the same reasons explained before.

\begin{figure}[H]
\centering
\begin{tabular}{cc}
\scalebox{0.22}{\includegraphics{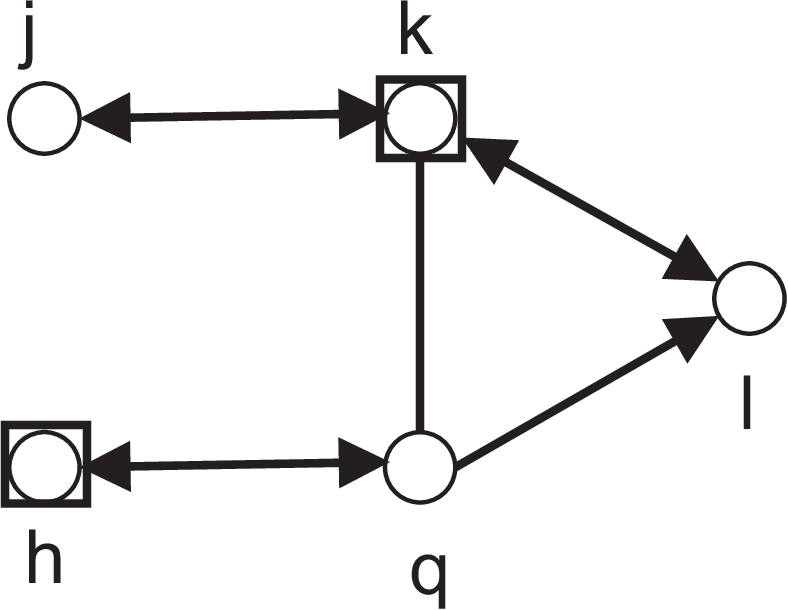}} \nn\nn\nn&\nn\nn\nn
\scalebox{0.22}{\includegraphics{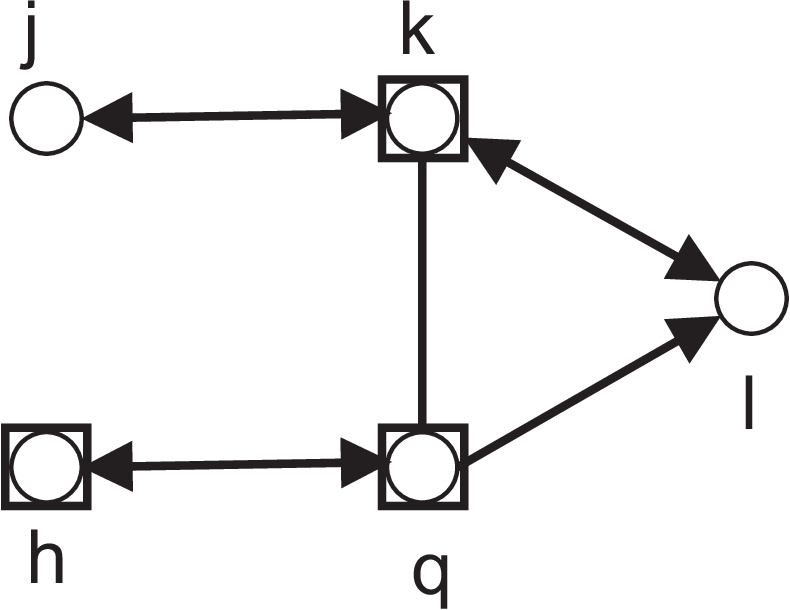}}\\
(a)\nn\nn\nn & \nn\nn\nn (b)\\
\scalebox{0.22}{\includegraphics{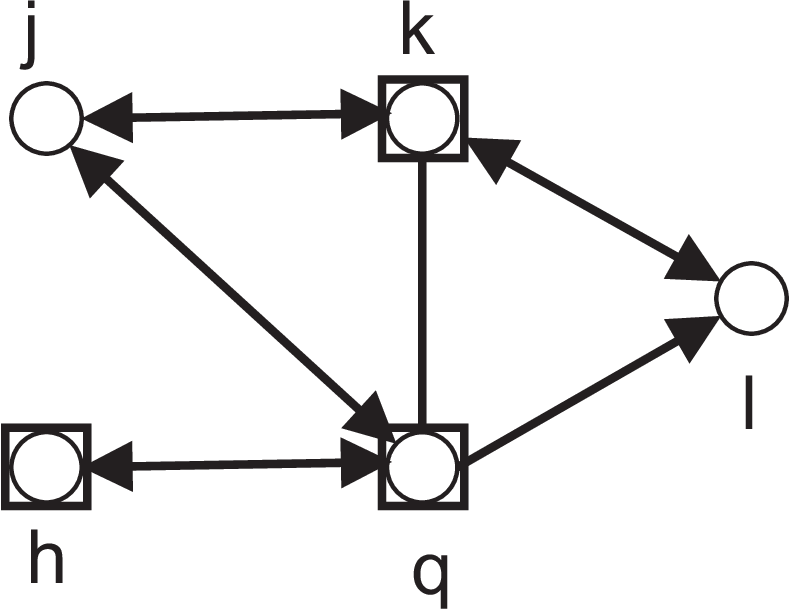}} \nn\nn\nn&\nn\nn\nn
\scalebox{0.22}{\includegraphics{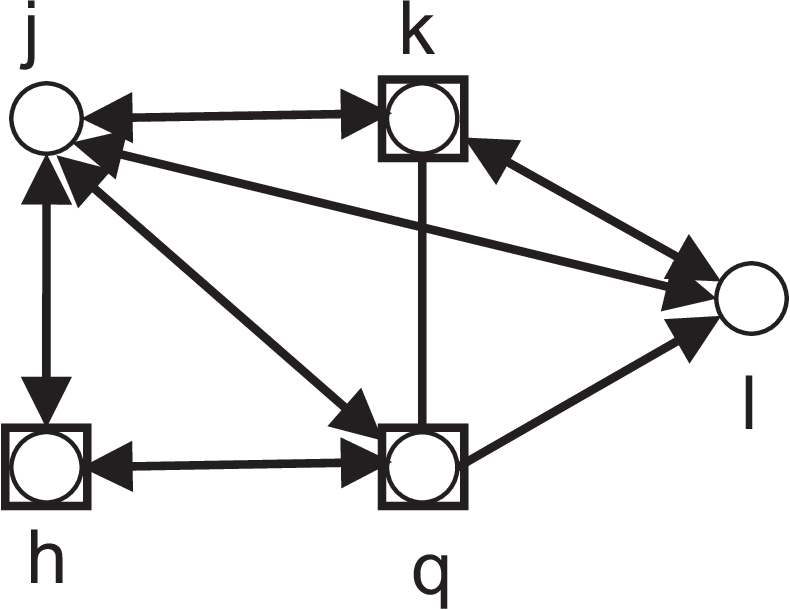}}\\
(c)\nn\nn\nn &\nn\nn\nn (d)\\
\scalebox{0.22}{\includegraphics{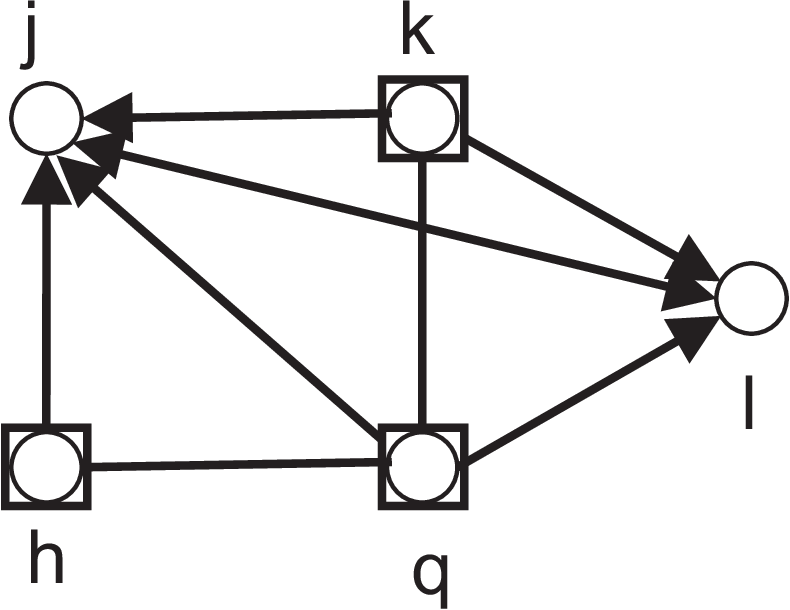}} \nn\nn\nn&\nn\nn\nn
\scalebox{0.22}{\includegraphics{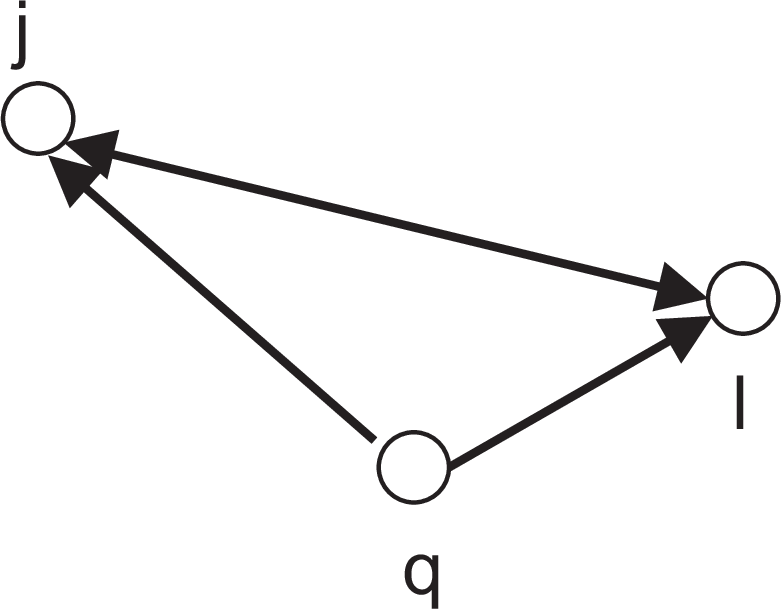}}\\
(e) \nn\nn\nn&\nn\nn\nn (f)
\end{tabular}
  \caption[]{\small{(a) A chain mixed graph $G$, $\condnc \in C$. (b) The graph after applying step 1 of Algorithm \ref{alg:3}, $\condnc \in S=C\cup \ant(C)$. (c) The generated graph after applying step 2 (step 5 of Table \ref{tab:2}).
  (d) The generated graph after applying step 3 (steps 2 and 3 of Table \ref{tab:2}). (e) The generated graph after applying step 4. (f) The generated CMG from $G$.}}
     \label{fig:alg3ex}
\end{figure}

Fig.\ \ref{fig:alg3ex} illustrates how to apply Algorithm \ref{alg:3} step by step to a CMG.
First, let us provide a global interpretation of step 3 of Algorithm \ref{alg:3}.
\begin{lemma}\label{lem:2}
Let $G$ be a CMG. There exists an edge between $i$ and $j$ in the graph generated after step 3 of Algorithm \ref{alg:3} if and only if there
exists an endpoint-identical walk to the edge between $i$ and $j$ in the generated graph after step 2 whose inner sections are all collider and in $C\cup \ant(C)$, and whose endpoint sections contain a single node ($i$ or $j$).
\end{lemma}
We provide two lemmas that explain why the set $S$ can be fixed in the beginning of the algorithm, and why there is no need to apply step 4 of Algorithm \ref{alg:3} repeatedly.
\begin{lemma}\label{lem:20}
Let $G$ be a CMG. If there is an arrow from $j$ to $i$ or a line between $j$ and $i$ generated by steps 3 or 4 of  Algorithm \ref{alg:3} then $j\in S=C\cup\ant(C)$. In addition, generated lines by Algorithm \ref{alg:3} do not lie on any collider section in $\alpha_{CMG}(G;\varnothing,C)$.
\end{lemma}
\begin{lemma}\label{lem:21}
Let $G$ be a CMG. A node $i$ is in $\ant(C)$ in $G$ if and only if it is in $\ant(C)$ in the graph generated after every step of  Algorithm \ref{alg:3} before step 5.
\end{lemma}
We now follow the same procedure as in the previous section.
\begin{prop}\label{prop:3}
Graphs generated by Algorithm \ref{alg:3} are CMGs.
\end{prop}
 Here, we provide the global interpretation of Algorithm  \ref{alg:3}.
 \begin{lemma}\label{lem:2n}
Let $G$ be a CMG. There exists an edge between $i$ and $j$ in $\alpha_{CMG}(G;\varnothing,C)$ if and only if there
exists a walk between $i$ and $j$ in $G$ whose inner sections are all collider and in $S=C\cup \ant(C)$, and whose endpoint sections contain a single node ($i$ or $j$) except when there is an arrowhead at the section containing $i$ (or $j$), and $i$ (or $j$) is a spouse of a member of $S$. In addition, the walk and the edge are endpoint-identical except when  there is an arrowhead at the endpoint section containing $i$ (or $j$), and $i\in \ant(C)$ (or $j\in \ant(C)$) in $G$.
\end{lemma}
\begin{theorem}\label{thm:3n}
For a chain mixed graph $G$ and disjoint subsets $C$ and $C_1$ of its node set,
\begin{displaymath}
\alpha_{CMG}(\alpha_{CMG}(G;\varnothing,C);\varnothing,C_1)=\alpha_{CMG}(G;\varnothing,C\cup C_1).
\end{displaymath}
\end{theorem}
\begin{theorem}\label{thm:3}
For a chain mixed graph $G$ and disjoint subsets $A$, $B$, $C$, and $C_1$ of its node set,
\begin{displaymath}
\langle A,B\cd C_1\rangle\in\mathcal{J}_c(\alpha_{CMG}(G;\varnothing,C)) \iff \langle A,B\cd C\cup C_1\rangle\in\mathcal{J}_c(G).
\end{displaymath}
\end{theorem}
\begin{coro}\label{coro:3}
The class of chain mixed graphs, $\mathcal{CMG}$, with $c$-separation criterion is stable under conditioning.
\end{coro}
Applying Algorithm \ref{alg:3} to a CG, step 2 becomes inapplicable, and step 3 specializes to generating a line between the endpoints of collider trislides with at least one inner node in $S$ if the line does not already exist. Denote this specialization by $\alpha_{CG}(G,\varnothing,C)$. We first have the following:
\begin{prop}\label{prop:1}
Algorithm \ref{alg:3} generates CGs from CGs.
\end{prop}
Denote now by $\mathcal{CG}$ the set of all CGs. We also provide the following trivial statement:
\begin{prop}\label{prop:1vn}
The map $\alpha_{CG}(\cdot;\varnothing,\cdot)$ from $\mathcal{CG}$ and a subset of the node set of its members to $\mathcal{CG}$ is surjective.
\end{prop}
\begin{proof}
The result follows from the fact that $\alpha_{CMG}(G;\varnothing,\varnothing)=G$.
\end{proof}
We, therefore, have the following immediate corollary:
\begin{coro}
The class of chain graphs, $\mathcal{CG}$, with the LWF Markov property is stable under conditioning.
\end{coro}

\subsection{Simultaneous marginalization and conditioning for CMGs}\label{sec:6.2} 
\begin{figure}[H]
\centering
\scalebox{0.25}{\includegraphics{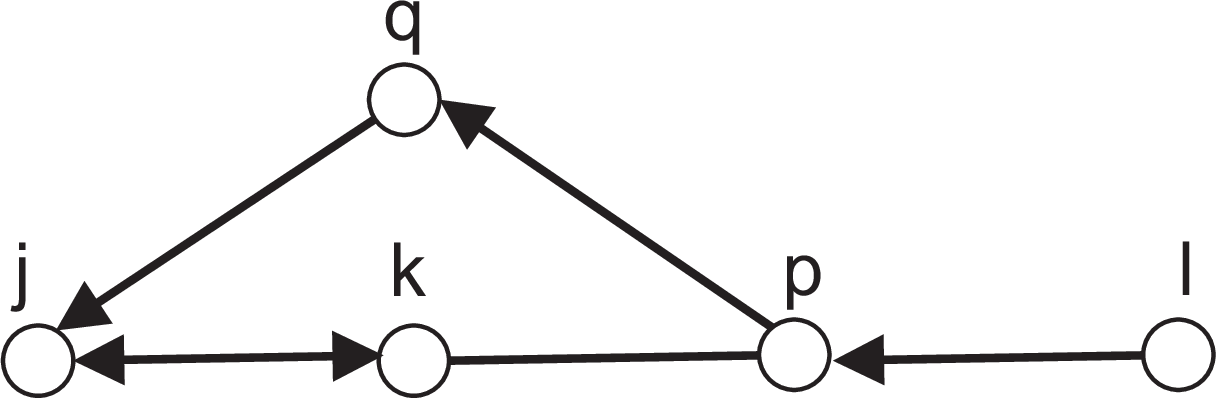}}
  \caption[]{\small{A non-maximal AnG.}}
     \label{fig:nonmaxex}
\end{figure}

Corollaries \ref{coro:2} and \ref{coro:3} imply that $\mathcal{CMG}$ with $c$-separation criterion is \emph{stable under marginalization and conditioning}, which formally holds when there is a graph $H\in\mathcal{CMG}$ such that $\mathcal{J}_c(H)=\alpha(\mathcal{J}_c(G);M,C)$, where
\begin{displaymath}
\alpha(\mathcal{J};M,C)=\{\langle A,B\cd D\rangle:\langle A,B\cd D\cup C\rangle\in\mathcal{J}\text{ and }(A\cup B\cup D)\cap(M\cup
C)=\varnothing\}.
\end{displaymath}
We now deal with the case where there are both marginalization and conditioning subsets in a CMG.  We first define maximality in order to simplify the results.
A graph is \emph{maximal} if to every non-adjacent pairs of nodes, there is an independence statement associated. CMGs are not maximal since, for example, the class of ancestral graphs \citep{ric02} is a subclass of CMGs, and there exist non-maximal ancestral graphs; see also Fig.\ \ref{fig:nonmaxex}, for an example of a CMG that is not ancestral and that induces no independence statement of form $j\dse_c l \cd C$ for any choice of $C$. There is a method to generate, from a non-maximal CMG, a maximal CMG that induces the same independence model, which is beyond the scope of this manuscript. However, here we provide a sufficient condition for non-maximal graphs as a lemma, which will be used in our proofs.

\begin{lemma}\label{lem:vvn}
If there is a collider trislide between $i$ and $j$ in $G$ such that there is an arrow from an inner node of the trislide to $j$ (or $i$) and $i\not\sim j$ then $G$ is not maximal.
\end{lemma}
We also provide the following lemma, which deals with the global behavior of the simultaneous marginalization and conditioning as described later in this section:
\begin{lemma}\label{lem:vu}
There is an edge between $i$ and $j$ in $\alpha_{CMG}(\alpha_{CMG}(G;\varnothing,C);M,\varnothing)$ if and only if there is a walk between $i$ and $j$ in $G$ on which (i) all nodes on collider sections are in $C\cup\ant(C)$; (ii) on non-collider sections, (a) all nodes  are in $M$, or (b) one endpoint is in $M$ and also either a child of a node in $M$ or a spouse of a node in $C\cup\ant(C)$, and the other endpoint has an arrowhead at it from the adjacent node on the walk. In addition, the walk and the edge are endpoint-identical except when there is an arrowhead at the endpoint section containing $i$ (or $j$), and $i\in \ant(C)$ (or $j\in \ant(C)$) in $G$.
\end{lemma}
We now have the following important result, which illustrates that, for maximal graphs, in order to both marginalize and condition, it does not matter whether we marginalize first by using Algorithm \ref{alg:2} and then condition by using Algorithm \ref{alg:3} or vice versa:
\begin{prop}\label{prop:l}
For a chain mixed graph $G$ and two disjoint subsets $M$ and $C$ of its node set, it holds that
\begin{displaymath}
\alpha_{CMG}(\alpha_{CMG}(G;M,\varnothing);\varnothing,C)=\alpha_{CMG}(\alpha_{CMG}(G;\varnothing,C);M,\varnothing)
\end{displaymath}
if $\alpha_{CMG}(\alpha_{CMG}(G;M,\varnothing);\varnothing,C)$ is maximal.
\end{prop}
It is also clear from the proof that if we drop the maximality assumption then the two concerned graphs in the proposition induce the same independence models. In addition, we show that the corresponding algorithm (Algorithm \ref{alg:2} followed by Algorithm \ref{alg:3} or vice versa) is well-defined for maximal graphs. We denote the corresponding function by $\alpha_{CMG}(G;M,C)$. In general, one can first apply Algorithm \ref{alg:3} followed by Algorithm \ref{alg:2}, in which case we showed in the proof that an edge is present between the endpoints of  the walk described in Lemma \ref{lem:vvn}.
\begin{theorem}\label{thm:4n}
For a chain mixed graph $G$ and disjoint subsets $M$, $M_1$, $C$, and $C_1$ of its node set,
\begin{displaymath}
\alpha_{CMG}(\alpha_{CMG}(G;M,C);M_1,C_1)=\alpha_{CMG}(G;M\cup M_1,C\cup C_1)
\end{displaymath}
if the two graphs are maximal.
\end{theorem}
\begin{proof}
The result follows from the definition and Proposition \ref{prop:l}, Theorem \ref{thm:3n}, and Theorem \ref{thm:2n}.
\end{proof}
In Proposition \ref{prop:2vn}, we showed that  all CGs after marginalization are mapped onto $\mathcal{H}$, which is a subclass of CMGs.  Here we show that CGs after marginalization and conditioning are also mapped onto $\mathcal{H}$.
\begin{prop}\label{prop:2vnn}
The map $\alpha_{CMG}$ maps $\mathcal{CG}$ and two subsets of the node set of its members surjectively onto $\mathcal{H}$.
\end{prop}
We are now ready to provide the main result, which illustrates that by applying Algorithm \ref{alg:2} followed by Algorithm \ref{alg:3} (or vice versa), we obtain the marginal and conditional independence model for a CMG (or a CG) after marginalization and conditioning.
\begin{theorem}\label{thm:4}
For a chain mixed graph $G$ and disjoint subsets $A$, $B$, $M$, $C$, and $C_1$ of its node set,
\begin{displaymath}
\langle A,B\cd C_1\rangle\in\mathcal{J}_c(\alpha_{CMG}(G;M,C)) \iff \langle A,B\cd C\cup C_1\rangle\in\mathcal{J}_c(G).
\end{displaymath}
\end{theorem}
\begin{proof}
By definition and Proposition \ref{prop:l}, Theorem \ref{thm:3}, and Theorem \ref{thm:2}, it is implied that
\begin{equation*}
\begin{split}
\langle A,B\cd C_1\rangle\in\mathcal{J}_c(\alpha_{CMG}(G;M,C))=\mathcal{J}_c(\alpha_{CMG}(\alpha_{CMG}(G;M,\varnothing);\varnothing,C)) \iff \\
\langle A,B\cd C\cup C_1\rangle\in \mathcal{J}_c(\alpha_{CMG}(G;M,\varnothing))\iff \langle A,B\cd C\cup C_1\rangle\in \mathcal{J}_c(G).
\end{split}
\end{equation*}
\end{proof}
\section{Anterial graphs}\label{sec:7}
The definition of CMGs can be considered a generalization of the definition of \emph{summary graphs} (SGs) by \cite{wer11}: CMGs collapse to SGs when there are no arrowheads pointing to lines. CMGs are also analogous to SGs in the sense that they capture the marginal and conditional models for CGs, and SGs capture the marginal and conditional models for DAGs; and CMGs exclude graphs with semi-directed cycles while SGs exclude graphs with directed cycles.

The class of ancestral graphs, defined by \cite{ric02}, captures the same independence models as those of SGs, but has a simpler structure than SGs. In this section, we define the class of \emph{anterial graphs} (AnGs), which can be thought of as a generalization of and analogous to ancestral graphs with the same relationship to CMGs as that of ancestral graphs to SGs.

\begin{figure}[H]
\centering
\begin{tabular}{cc}
\scalebox{0.15}{\includegraphics{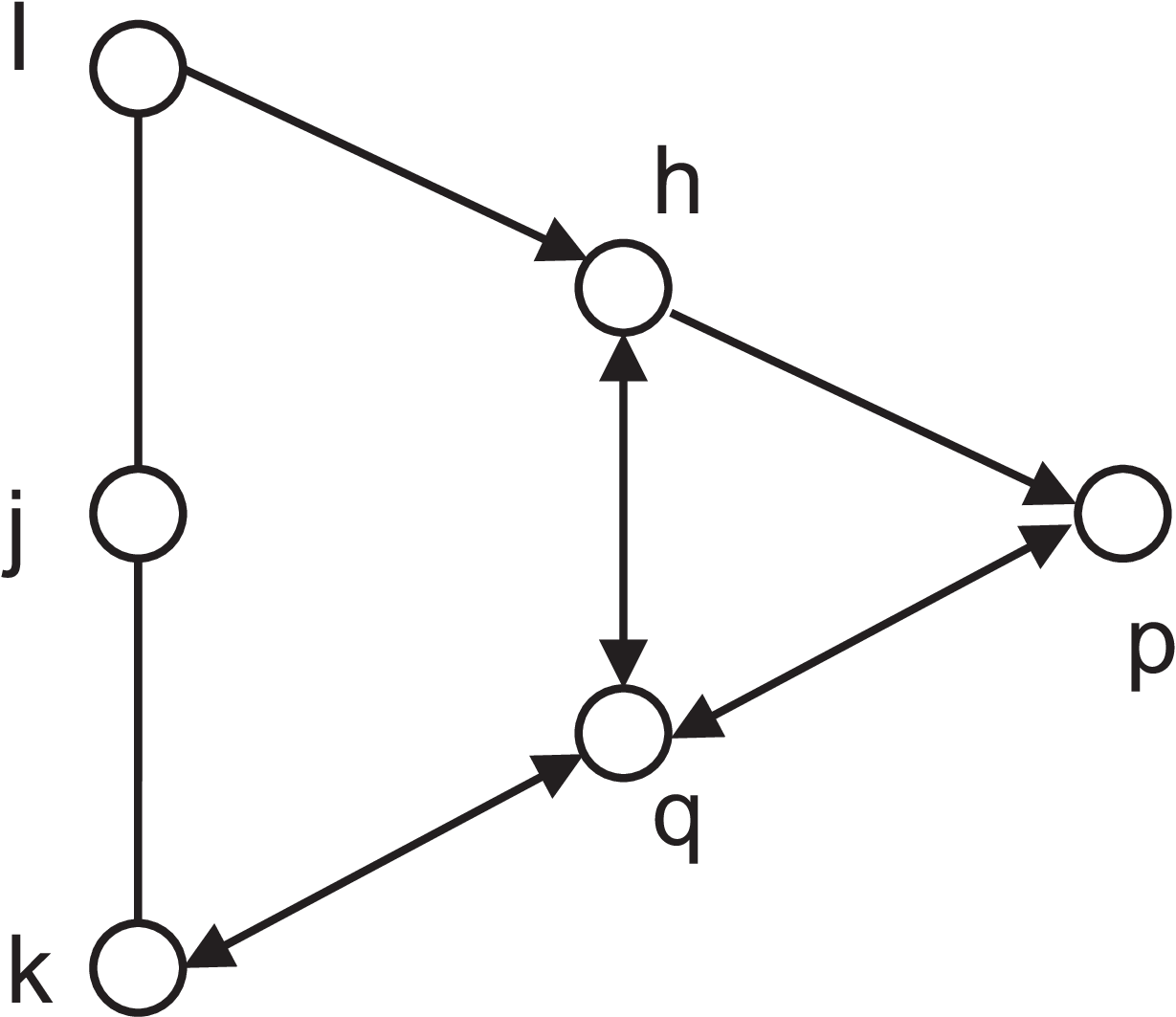}}\nn\nn &
\nn\nn\scalebox{0.15}{\includegraphics{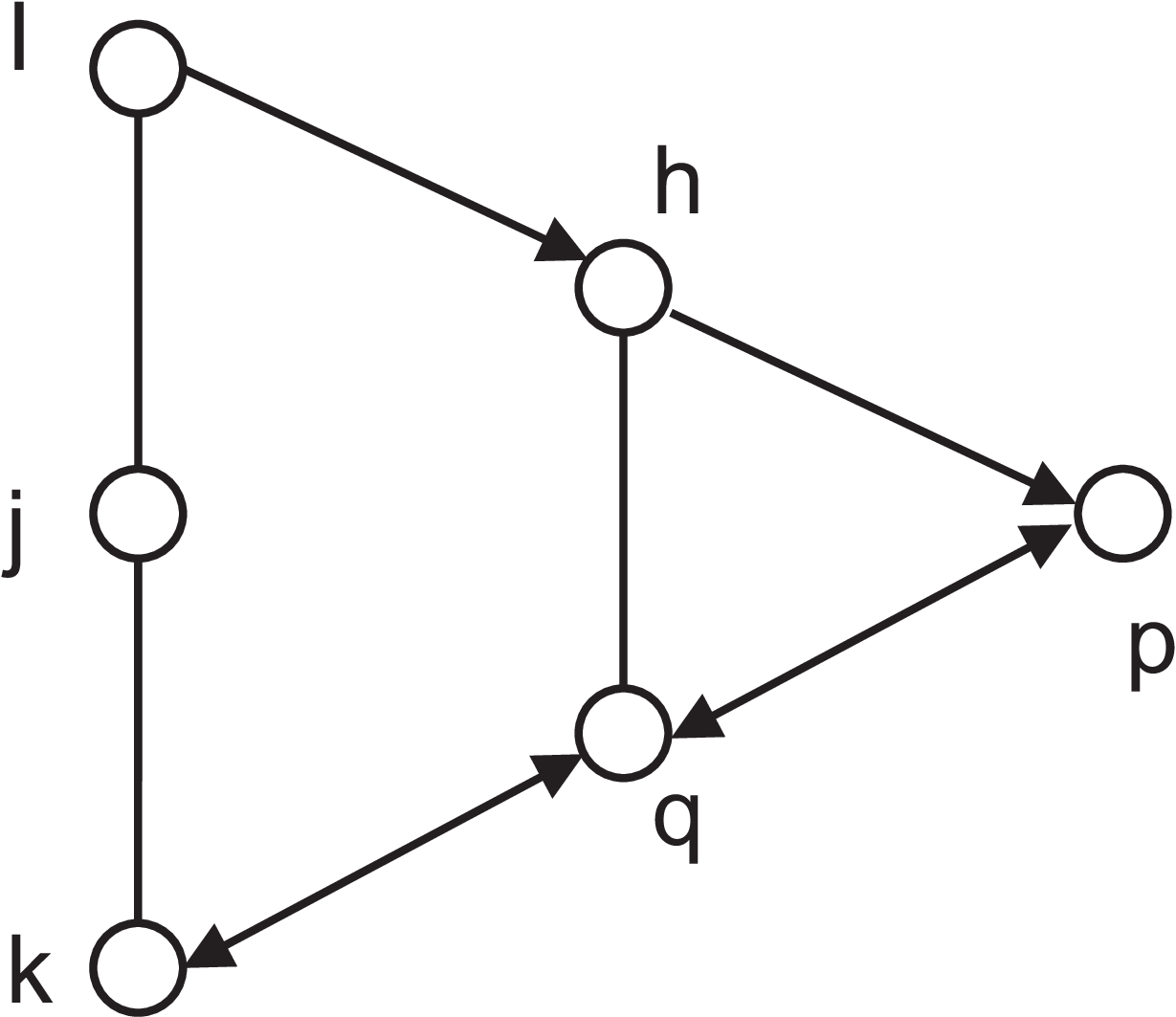}}\\
(a) & (b)
\end{tabular}
  \caption[]{\small{(a) An AnG. (b) A CMG that is not an AnG.}}
     \label{fig:AnGex}
\end{figure}

An anterial graph is a mixed graph that contains neither semi-directed cycles that contain at least an arrow; nor does it contain arcs with one endpoint that is an anterior of the other endpoint. This implies that, unlike CMGs, AnGs are simple graphs. For example, in Fig.\ \ref{fig:AnGex}(a) the graph is an AnG, but in Fig.\ \ref{fig:AnGex}(b) the graph is not an AnG because of the existence of the arc $kq$, where $k\in\ant(q)$ via the semi-directed path $\langle k,j,l,h,q\rangle$ as well as the arc $qp$, where $q\in\ant(p)$.

Here we show that, from an anterial graph and after marginalization and conditioning, how to generate an anterial graph with the corresponding marginal and conditional independence model.
\begin{alg}\label{alg:4}
\text{$\alpha_{AnG}(G;M,C)$: (Generating an AnG from an anterial graph $G$)}\\
Start from $G$.
\begin{enumerate}
 \item Apply Algorithm \ref{alg:3}.
 \item Apply Algorithm \ref{alg:2}.
 \item Generate respectively arrows from $j$ to $i$ or arcs between $i$ and $j$ for trislides $j\fra \circ\ful\cdots\ful i\arc k$ or $j\arc \circ\ful\cdots\ful i\arc k$ when $k\in\ant(i)$ if the arrow or the arc does not already exist.
 \item Generate respectively an arrow from $j$ to $i$ or an arc between $i$ and $j$ for trislides $j\fra k_1\ful\cdots\ful k_m\arc i$ or $j\arc k_1\ful\cdots\ful k_m\arc i$ when there is an $1\leq r\leq m$ such that $k_r\in\ant(i)$ if the arrow or the arc does not already exist. Continually apply this step until it is not possible to apply it further.
    \item Remove the arc between $j$ and $i$ in the case that $j\in \ant(i)$, and replace it with an arrow from $j$
    to $i$ if the arrow does not already exist; and remove the arc between $j$ and $i$ in the case that $j\in \ant(i)$ and $i\in\ant(j)$, and replace it with a line between $i$
    and $j$ if the line does not already exist.
\end{enumerate}
\end{alg}
 Notice that, as we will see, steps 3, 4, and 5 of Algorithm \ref{alg:4} generate, from the generated CMG after step 2, an AnG that captures the same independence model as that of the CMG. In addition, in step 4, one $k_r$ being in $\ant(i)$ implies that all $k_r$, $1\leq r\leq m$, are in $\ant(i)$, and in this sense we can say that a section is in $\ant(i)$.

This Algorithm is a generalization of the related algorithm for ancestral graphs \cite{ric02,sad13}. Again, one can see that sections here are treated in the same way as nodes in the ancestral-graph-generating algorithms. The idea here is that step 4 generates a dependency between $j$ and $i$ (which in fct always exists) before step 5 makes the graph anterial, and consequently destroys the dependency between $i$ and $j$.

\begin{figure}
\centering
\begin{tabular}{cc}
\scalebox{0.22}{\includegraphics{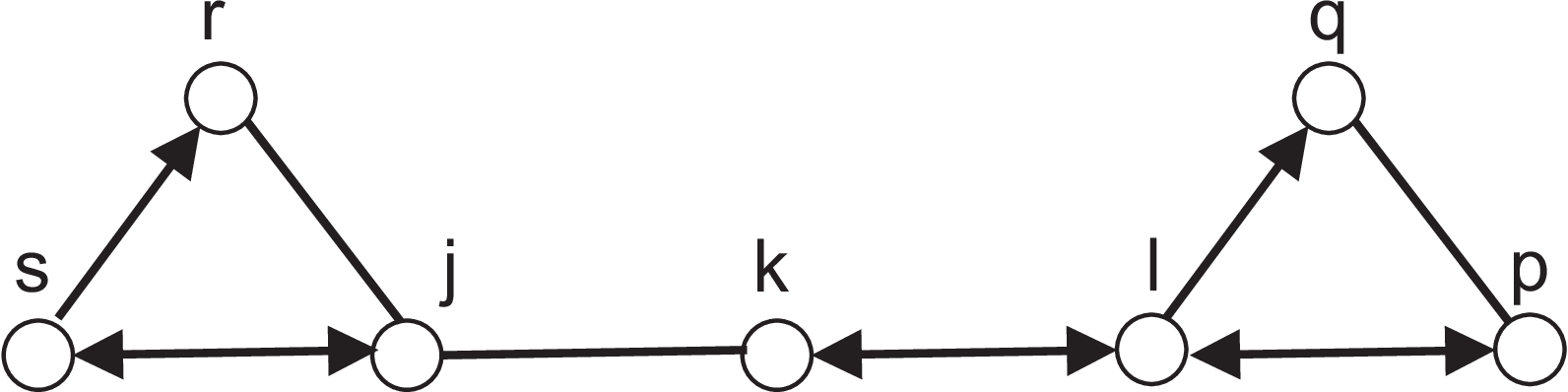}} &
\scalebox{0.22}{\includegraphics{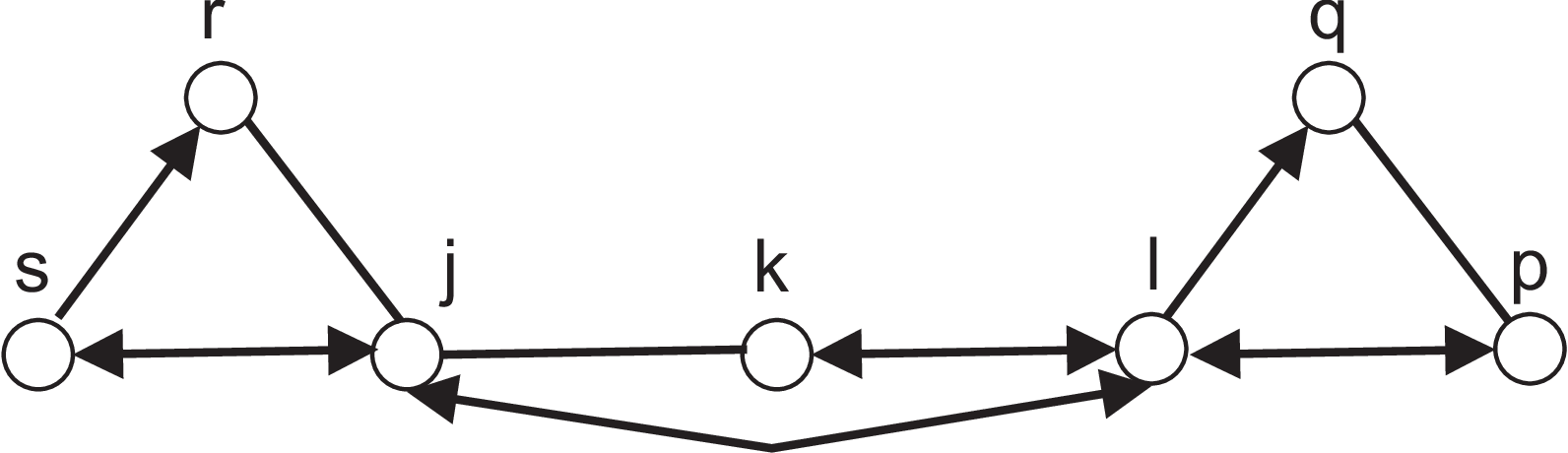}}\\
(a) & (b)\\
\scalebox{0.22}{\includegraphics{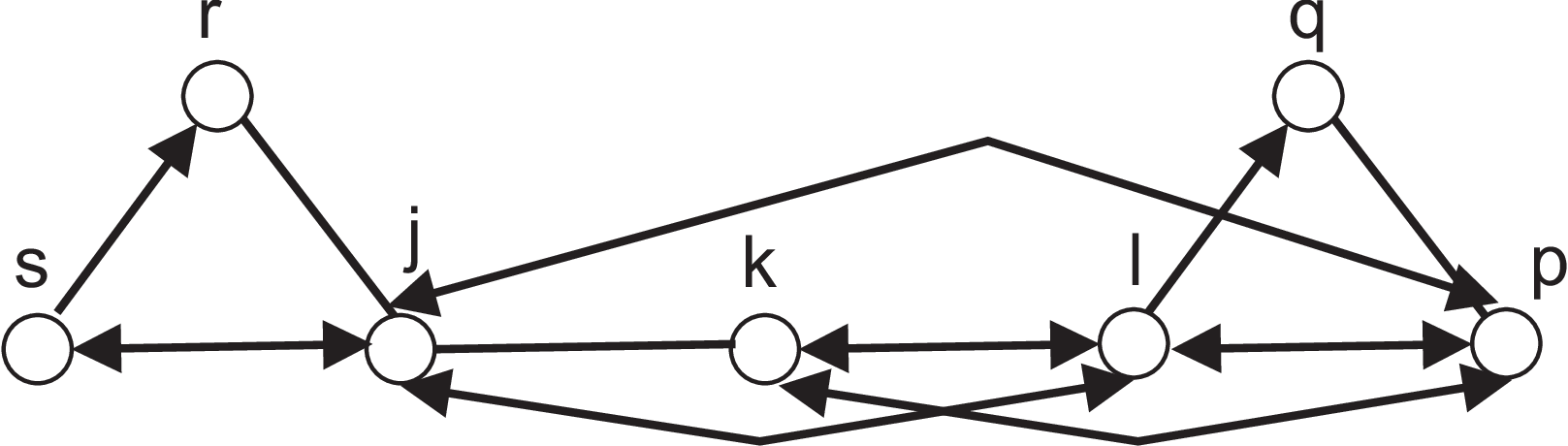}} &
\scalebox{0.22}{\includegraphics{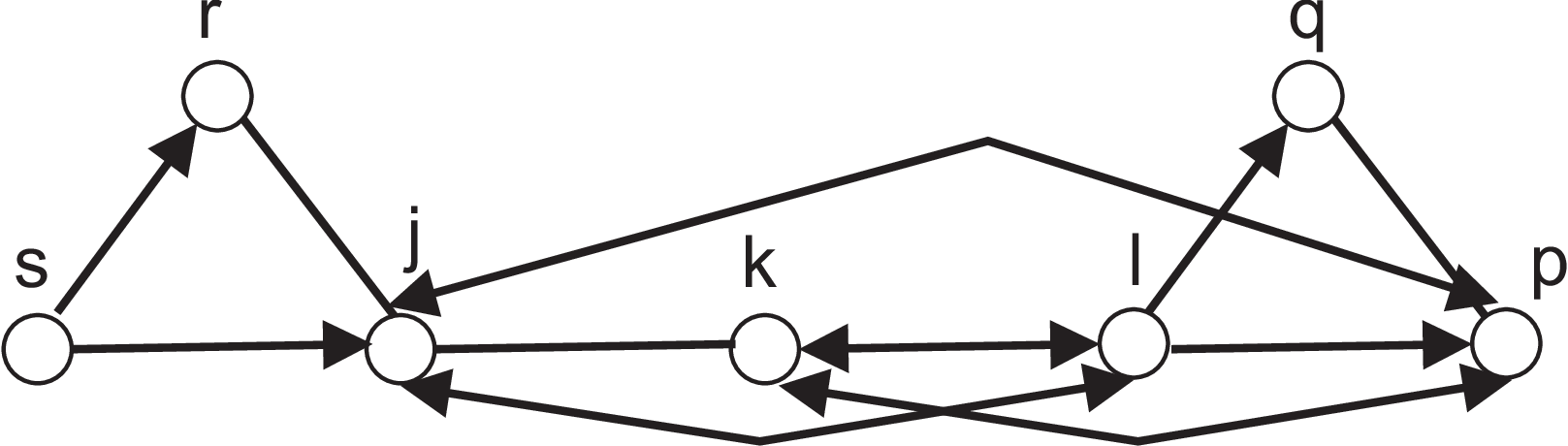}} \\
(c) & (d)
\end{tabular}
  \caption[]{\small{(a) A chain mixed graph $G$. (b) The graph after applying step 3 of Algorithm \ref{alg:4}. (c) The graph after applying step 4.
  (d) The generated AnG after applying step 5.}}
     \label{fig:alg4ex}
\end{figure}

Fig.\ \ref{fig:alg4ex} illustrates how to apply  these steps to a CMG.
We consider Algorithm \ref{alg:4} a function denoted by $\alpha_{AnG}$. Notice that for every anterial graph $G$, it holds that $\alpha_{AnG}(G;\varnothing,\varnothing)=G$. We again follow a parallel theory as that in the previous sections:
\begin{prop}\label{prop:5}
Graphs generated by Algorithm \ref{alg:4} are AnGs.
\end{prop}
We first provide two lemmas that deal with the global behavior of the algorithm.
\begin{lemma}\label{lem:30}
Let $H$ be a chain mixed graph. It holds that $i\in\ant(j)$ in $H$ if and only if $i\in\ant(j)$ in the anterial graph generated after applying steps 3, 4, and 5 of Algorithm \ref{alg:4} to $H$.
\end{lemma}
Denote by a walk between $i$ and $j$ on which all sections are collider and every inner section is in $\ant(i)$ a \emph{subprimitive inducing walk} from $j$ to $i$. This is a  special case of a generalization of primitive inducing paths, defined in \cite{ric02}, where all nodes are anteriors of one of the endpoints, not either of the endpoints. We also denote the function corresponding to steps 3, 4, and 5 of Algorithm \ref{alg:4} by $\alpha_{CMG.AnG}$. Notice that $\alpha_{AnG}(G;M,C)= \alpha_{CMG.AnG}(\alpha_{CMG}(G;M,C))$.
\begin{lemma}\label{lem:31}
Let $H$ be a chain mixed graph. There is an edge between $i$  and $j$ in $\alpha_{CMG.AnG}(H)$
if and only if there is a subprimitive inducing walk from $j$ to $i$ in $H$ (which might also contain $i$ as an inner node) with single-element endpoint sections.
In addition, the edge and the walk are endpoint-identical except when $i\in\ant(j)$ or $j\in\ant(i)$ in $H$, in which case there is no arrowhead at $i$ or at $j$, respectively, on the $ij$ edge in $\alpha_{CMG.AnG}(H)$.
\end{lemma}
We now prove that Algorithm \ref{alg:4} does not need to be applied to an anterial graph, but it can be applied to a chain mixed graph.
\begin{lemma}\label{lem:3}
Let $H$ be a chain mixed graph and $M$ and $C$ be two subsets of its node set. It holds that $\alpha_{AnG}(\alpha_{CMG.AnG}(H);M,C)=\alpha_{AnG}(H;M,C)$.
\end{lemma}
\begin{theorem}\label{thm:5n}
For an anterial graph $G$ and disjoint subsets $M$, $M_1$, $C$, and $C_1$ of its node set,
\begin{displaymath}
\alpha_{AnG}(\alpha_{AnG}(G;M,C);M_1,C_1)=\alpha_{AnG}(G;M\cup M_1,C\cup C_1),
\end{displaymath}
if the two graphs are maximal.
\end{theorem}
\begin{proof}
Using Theorem \ref{thm:4n} and Lemma \ref{lem:3}, we have the following:
\begin{displaymath}
\alpha_{AnG}(\alpha_{AnG}(G;M,C);M_1,C_1)=\alpha_{AnG}(\alpha_{CMG.AnG}(\alpha_{CMG}(G;M,C));M_1,C_1)=
\end{displaymath}
\begin{displaymath}
\alpha_{AnG}(\alpha_{CMG}(G;M,C);M_1,C_1)=
\alpha_{CMG.AnG}(\alpha_{CMG}(\alpha_{CMG}(G;M,C);M_1,C_1))=
\end{displaymath}
\begin{displaymath}
\alpha_{CMG.AnG}(\alpha_{CMG}(G;M\cup M_1,C\cup C_1))=\alpha_{AnG}(G;M\cup M_1,C\cup C_1).
\end{displaymath}
\end{proof}
Denote the set of all AnGs by $\mathcal{ANG}$.
\begin{prop}\label{prop:5vn}
Let $\mathcal{K}$ be the subset of $\mathcal{ANG}$  with the following properties:
\begin{enumerate}
  \item There is no collider trislide of form $k\arc i\ful\dots\ful j\fla l$ unless there is an arrow from $l$ to $i$.
  \item There is no collider trislide of form $k\arc i\ful\dots\ful j\arc l$ unless there are $jk$ and $il$ arcs and an $ij$ line.
\end{enumerate}
Then $\alpha_{AnG}$ maps $\mathcal{CG}$ and two subsets of the node set of its members surjectively onto $\mathcal{K}$.
\end{prop}
\begin{theorem}\label{thm:5}
For an anterial graph $G$ and disjoint subsets $A$, $B$, $M$, $C$, and $C_1$ of its node set,
\begin{displaymath}
\langle A,B\cd C_1\rangle\in\mathcal{J}_c(\alpha_{AnG}(G;M,C)) \iff \langle A,B\cd C\cup C_1\rangle\in\mathcal{J}_c(G).
\end{displaymath}
\end{theorem}
\begin{coro}
The class of anterial graphs, $\mathcal{ANG}$, with $c$-separation criterion is stable under marginalization and conditioning.
\end{coro}\label{coro:5}

\section{Probabilistic independence models for CMGs and AnGs and comparison to other types of graphs}\label{sec:8}
The most interesting independence models are induced by probability distributions. Consider a set $V$ and a collection of random variables
$(X_\alpha)_{\alpha\in V}$ with joint density $f_V$. By letting $X_A=(X_v)_{v\in A}$ for each subset $A$ of $V$,
we then use the short notation $A\cip B\cd C$ for $X_A\cip X_B\cd X_C$ and
disjoint subsets $A$, $B$, and $C$ of $V$.

For a given independence model $\mathcal{J}$, a probability distribution $P$ is called \emph{faithful} with respect to $\mathcal{J}$ if, for random vectors $X_A$, $X_B$, and $X_C$ with probability distribution $P$,
\begin{displaymath}
A\cip B\cd C \text{ if and only if } \langle A,B\cd C\rangle\in \mathcal{J}.
\end{displaymath}
We say that $\mathcal{J}$ is \emph{probabilistic} if there is a distribution $P$ that is faithful to $\mathcal{J}$.

From a given collection of random variables $(X_\alpha)_{\alpha\in V}$ with a probability distribution $P$, one can induce an independence model $\mathcal{J}(P)$ by demanding
\begin{displaymath}
\text{if } A\cip B\cd C \text{ then } \langle A,B\cd C\rangle\in \mathcal{J}(P).
\end{displaymath}
Notice that $\mathcal{J}(P)$ is obviously probabilistic.

For a chain graph $G$, we say that a probability distribution with density $f$ factorizes with respect to $G$ if
\begin{displaymath}
f(x)=\prod_{\tau\in \mathcal{T}}f(x_{\tau}\cd x_{\pa(\tau)}),
\end{displaymath}
where $\mathcal{T}$ is the set of chain components of $G$; and
\begin{displaymath}
f(x_{\tau}\cd x_{\pa(\tau)})=\prod_{a}\phi_a(x),
\end{displaymath}
where $a$ varies over all subsets of $\tau\cup\pa(\tau)$ that are complete in the moral graph of the subgraph of $G$ induced by $\tau\cup\pa(\tau)$, and $\phi_a(x)$ is a function that depends on $x$ through $x_a$ only; see \cite{lau96} for more discussion.

Now let $\alpha(P,;M,C)$ be the probability distribution obtained by usual probabilistic marginalization and conditioning for the probability distribution $P$. It is easy to show that if $P$ is faithful to  $\mathcal{J}$ then $\alpha(P,;M,C)$ is faithful to the marginal and conditional independence model $\alpha(\mathcal{J};M,C)$; see Theorem 7.1 and Corollary 7.3 of \cite{ric02}.

It is also known that if $G$ is a CG then there is a regular Gaussian distribution that is faithful to it. In fact, almost all the regular Gaussian
distributions that factorize with respect to a CG are faithful to it; see \cite{pen11}. In other words, the independence mode $\mathcal{J}_c(G)$ is probabilistic.

By Propositions  \ref{prop:2vn}, \ref{prop:2vnn}, and \ref{prop:5vn}, a considerably large subclass of CMGs or AnGs are obtained by chain graphs after marginalization and conditioning. Hence, it is implied by the discussion above that for a graph $H$ in these subclasses, $\mathcal{J}_c(H)$ is probabilistic; i.e.\ there is a distribution (in fact at least a Gaussian distribution) that is faithful to it.


One can obtain the same result for the strictly positive discrete probability distributions since there is such a distribution that is faithful to a given CG \cite{stub98}. These results motivate the use of CMGs and AnGs.

The next, and probably more important, question in order to justify the use of these classes is whether it is possible to find a parametrization, e.g.\ Gaussian or discrete, of these graphs.

In the Gaussian case, there exists a known parametrization for the regular Gaussian distributions that factorize with respect to a CG; see \cite{wer06} and \cite{pen11} for two slightly different but equivalent parametrizations. For maximal ancestral graphs (MAGs), there is a known parametrization in the Gaussian \cite{ric02}. We believe that it is possible to extend this parametrization to the class of maximal AnGs. Here is some possible actions in order to generalize this parametrization.

Notice first that the classes of CMGs and AnGs are not maximal, as explained in Section \ref{sec:6.2}.  However, as mentioned before, there is a method to generate, from non-maximal CMGs and AnGs, maximal CMGs and AnGs that induce the same independence models. Hence, one can then focus on the class of maximal AnGs.

Considering the Gaussian parametrization for MAGs, one then needs to define, instead of one matrix for the undirected part of the MAG, one symmetric matrix for every chain component of the maximal AnG (as it is done in the Gaussian parametrization for CGs). It is also needed to generalize the ordering associated to MAGs, e.g.\, by defining an ordering for chain components containing lines instead of an ordering for the nodes. One may then follow the method described in Section 8 of the mentioned paper.

Since both parametrizations for CGs and MAGs are curved exponential families, and consequently the models associated with them are identifiable, the generalization for AnGs seems to preserve this desirable property.

Introducing a discrete parametrization for CMGs or AnGs seems much trickier. Similar to the Gaussian case, the goal should be to find a combination of discrete parametrizations for CGs (see, e.g\ \cite{pen09}) and summary graphs (or alternatively ADMGs -- see \cite{eva14}). For CMGs, a parametrization may be derived from the original CG with the use of structural equation models with latent variables. This can be considered a generalization of the method utilized in summary graph models.

Nonetheless, we again stress the importance of introducing different smooth parametrizations for CMGs and AnGs in a future work as well as studying additional non-independence constraints that arise in such models.

Besides the relevant parametrizations, it is clear that CMGs act similarly to summary graphs in the problem of marginalization and conditioning for DAGs, and AnGs act similarly to ancestral graphs. To give a more detailed comparison between CMGs (and AnGs) and summary graphs (and ancestral graphs), we first note that the lines in all these graphs have the same meaning. As mentioned before, there are no arrowheads at lines in the latter types, and one can think of sections with arrowheads pointing to them in the former types in the same manner as the nodes in the latter types. Indeed summary graphs and ancestral graphs are subclasses of CMGs and AnGs respectively, thus every summary or ancestral graph model is a CMG or AnG model.

In addition, in CMGs, for a collider trislide of from $i\fra j\ful l\fla k$, it holds that $i\notdse_c l$, $i\notdse_c l\cd j$, but $i\dse_c l\cd\{j,k\}$. However, there is no summary graph that can capture the same independencies and dependencies. Hence, for any induced path with $4$ nodes (and, of course, for longer paths), one can provide a CMG that is associated to a different model than summary graph models. By this, it is clear that the class of CMG models is rich in the sense that when the number of nodes grows, the number of distinct CMG models grows faster than the number of distinct summary graph models.

The class of marginal AMP chain graphs (MAMP CGs) deals with a similar problem of marginalization for AMP chain graphs. The lines in these graphs have a different meaning in independence interpretation (they are related to lines in AMP CGs), and naturally the class of models they represent is quite different. However, both classes of models contain the class of regression graph models \cite{wers11}, which itself contains the classes of undirected (concentration) graph models and the class of multivariate regression chain graph models as a subclass. In fact, if in a CMG, there is a section with non-adjacent endpoints that is larger than a single node then it can be seen that no MAMP CG can induce the same independence statements. This implies that, in the intersection of CMG and MAMP CG models, there is no arrowhead pointing to lines (in CMG sense). Therefore, this intersection is the same as the intersection of maximal ancestral graph and MAMP CG models (since MAMP CGs are maximal, and maximal summary and ancestral graphs induce the same independence model).
\section*{Acknowledgements}
The author is grateful to Steffen Lauritzen and Thomas Richardson for helpful discussions, Nanny Wermuth for helpful discussions and comments, and anonymous referees for the most helpful comments, especially detecting an error in the results.
\section*{Appendix: proofs}
In the Appendix, we provide proofs of the non-trivial lemmas, propositions, and theorems as well as some more technical and yet less informative lemmas that are used in the proofs.
\begin{proof}[Proof of Lemma \ref{lem:0}]
{\bf ($\Rightarrow$)} Suppose that there is a $c$-connecting walk $\pi$ between $i$ and $j$ given $C$.  Consider the shortest subpath $\rho_0$ of the section $\rho$ of $\pi$ between $k$ and $l$. If $\rho$ is a collider then a node of $\rho$ is in $C$, and since all the nodes on $\rho$ (including those on $\rho_0$) are connected by lines, they are all in $C\cup\ant(C)$. If $\rho$ is a non-collider then  all the nodes on $\rho$ (including those on $\rho_0$) are outside $C$. Hence, by replacing all such $\rho$ by $\rho_0$ we obtain the desired walk.

{\bf ($\Leftarrow$)} Suppose that there is a walk $\pi$ between $i$ and $j$ whose sections are all paths and nodes of every collider section are in $C\cup\ant(C)$, and non-collider sections are outside $C$. We keep all non-collider sections of $\pi$ intact. For a collider section $\rho$ between $k$ and $l$, if there is a node of $\rho$ in $C$, we keep it intact. Otherwise we replace $\rho$ with $\rho_4=\langle k,\rho_1,\rho_2,c,\rho_2^r,\rho_3,l\rangle$, where $\rho_1$ is a subpath of $\rho$ between $k$ and $h$ , $\rho_2$ is a semi-directed path from $h$ to a member $c$ of $C$, $\rho_2^r$ is $\rho_2$ in the reverse direction, and $\rho_3$ is a subpath of $\rho$ between $h$ and $l$. It is easy to observe that $\rho_4$ is $c$-connecting given $C$. (If there is an arrow on $\rho_2$ then $\rho_4$ consists of non-collider sections containing $\rho_1$ and $\rho_3$, and a collider section containing $c$; otherwise $\rho_4$ is a collider section containing $c$.) In addition, $\rho$ and $\rho_4$ are endpoint-identical. Hence, by this replacement for all such $\rho$ on $\pi$, we obtain a $c$-connecting walk given $C$ between $i$ and $j$.

Finally, from the construction of walks that we have in both directions of the proof, it is seen that the walks are endpoint-identical.
\end{proof}
\begin{proof}[Proof of Proposition \ref{prop:2}]
The resulting graphs have obviously the three desired types of edges, thus it is enough to prove that there is no semi-directed cycle that contains an arrow in the graph. Suppose, for contradiction, that there exists such a cycle. It is easy to observe that by replacing a generated line or arrow with the generating tripaths (cases 1, 2, 6, and 7 of Table \ref{tab:1}) or trislide (case 8), a semi-directed path remains semi-directed. Therefore, it is implied inductively that there is a semi-directed path in the original chain graph. This also contains an arrow since an arrow can only be replaced by a tripath or a trislide that contains an arrow. This is a contradiction.
\end{proof}
\begin{proof}[Proof of Lemma \ref{lem:1}]
{\bf ($\Leftarrow$)} Suppose that there exists a walk $\pi$ between $i$ and $j$  in the graph generated after applying step 1 of Algorithm \ref{alg:2} to $G$ whose inner sections are all non-collider and whose inner nodes are all in $M$. By Algorithm \ref{alg:2}, for a section between $k$ and $l$, a line between $k$ and $l$ is generated, and then, for a tripath $\langle h, q, r\rangle$ consisting of a line $hq$ with $q\in M$, the same edge as $qr$ is generated. Therefore, a walk is generated between $i$ and $j$ whose inner nodes are in $M$, and on which lines may only be adjacent to $i$ and $j$, and every section is a non-collider. By applying steps of Table \ref{tab:1}, we trivially obtain an endpoint-identical edge between $i$ and $j$.

{\bf ($\Rightarrow$)} Suppose that there is an edge between $i$ and $j$ in $\alpha_{CMG}(G;M,\varnothing)$. We are only interested in the case where this edge does not exist after  applying step 1 of Algorithm \ref{alg:2}. In this case, this edge is generated by step 2 by one of the tripaths in steps $1$ to $7$ of Table \ref{tab:1} in an iteration of step 2. Each edge in the tripath may have now been generated by a tripath with the inner node in $M$. By an inductive argument, we imply that in  the graph generated after applying step 1 of Algorithm \ref{alg:2} to $G$, there is a walk $\pi$ (because of possible self-intersections) between $i$ and $j$ whose inner nodes are in $M$. We show that there is no collider section on $\pi$: If, for contradiction, there is a collider section $\rho$ with endpoints $\langle k,\rho,l\rangle$ then it is easy to observe that, in some iteration of the algorithm, we obtain a collider tripath with endpoints $k$ and $l$, but no edge can be generated between $k$ and $l$ by the algorithm. Hence, there is no edge between $i$ and $j$ in $\alpha_{CMG}(G;M,\varnothing)$, a contradiction. Since in every iteration of the algorithm, the existence of an arrowhead at sections containing $i$ and $j$ does not change, $\pi$ remains endpoint-identical to the $ij$ edge.
\end{proof}
\begin{lemma}\label{lem:10}
Let $G$ be a CMG and $M$ a subset of its node set. If there is a path $i\ful\cdots\ful k\fla\, j$ or  $i\ful\cdots\ful k\arc\, j$ in $G$, and there is a semi-directed path of form $m_1\fra m_2\ful\dots \ful m_r\ful i$ with $m_s\in M$, $1\leq s\leq r$ then  Algorithm \ref{alg:2} generates an arrow from $j$ to $i$ or an arc  between $i$ and $j$, respectively.
\end{lemma}
\begin{proof}
Consider the section between  $k$, $i$, and $m_2$. By step 1 of Algorithm \ref{alg:2}, an arrow from $j$ to $m_2$ or a $jm_2$ arc is generated. Now by Lemma \ref{lem:1}, when we apply step 2 of the algorithm, an arrow from $j$ to $i$ or an $ij$ arc is generated.
\end{proof}
\begin{lemma}\label{lem:11}
Let $G$ be a CMG and $M$ a subset of its node set. There is a walk $\pi$ in $G$ with sections $\{\rho_1,\dots,\rho_r\}$ if and only if there is an endpoint-identical walk $\pi'$ in the graph generated after applying step 1 of Algorithm \ref{alg:2} for $M$ with sections $\{\rho'_1,\dots,\rho'_r\}$ such that $\rho'_q$ is a subsection of $\rho_q$ for $1\leq q\leq r$.  In addition, every node on $\pi$ that is not on $\pi'$ is on a subsection of $\pi$ with endpoints $l$ and $k$ such that $l$ exists on $\pi'$ and is a child of a member of $M$, and there is an arrowhead to $k$ on $\pi$.
\end{lemma}
\begin{proof}
The result follows from the fact that by replacing arrows and arcs on $\pi'$ by paths in cases 8 and 9 of Table \ref{tab:1} (the replacements that have occurred in step 1 of Algorithm \ref{alg:2}), sections become larger and no new section is generated; and vice versa.
\end{proof}
\begin{proof}[Proof of Theorem \ref{thm:2n}]
{\bf ($\Rightarrow$)} Suppose that in $\alpha_{CMG}(\alpha_{CMG}(G;M,\varnothing);M_1,\varnothing)$, there is an edge between $i$ and $j$. Notice that $i,j\notin M\cup M_1$. We prove that there is the same edge in $\alpha_{CMG}(G;M\cup M_1,\varnothing)$. Starting from an edge between $i$ and $j$, we discuss the type of path or walk that exists between $i$ and $j$ in every graph generated by different steps of Algorithm \ref{alg:2}:

{\bf In the graph generated before applying step 2 of Algorithm \ref{alg:2} to $\alpha_{CMG}(G;M,\varnothing)$ for $M_1$:} By Lemma \ref{lem:1}, there exists an endpoint-identical walk $\pi$ between $i$ and $j$  whose inner sections are all non-collider and inner nodes are all in $M_1$.

{\bf In $\alpha_{CMG}(G;M,\varnothing)$:} By Lemma \ref{lem:11}, there is a new walk, denoted by $\pi_1$. Define $l$ also as defined in the lemma, and notice that in this case $l$ is both in $M_1$ and a child of $m_1\in M_1$.

{\bf In the graph generated before applying step 2 of Algorithm \ref{alg:2} to $G$ for $M$:}
For every edge of $\pi_1$, again by Lemma \ref{lem:1}, there
exists an endpoint-identical walk between its endpoints, but with inner nodes in $M$. Denote the new walk generated by replacing all edges of $\pi_1$ by endpoint-identical walks at this stage by $\pi_2$. Notice that, because of endpoint-identicality, all nodes on $\pi_1$ remain non-collider on $\pi_2$. In addition, the $m_1l$ arrow might turn into a walk that contains a subwalk of from $m\fra m_2\ful\dots \ful m_r\ful l$  with $m\in M\cup M_1$ and $m_s\in M$, $2\leq s\leq r$.

{\bf In $G$:}  Again by Lemma \ref{lem:11}, there is a new walk, denoted by $\pi_3$. Notice that the arrow from $m$ to $m_2$ might be replaced by a path, but nevertheless, by possibly changing the node $m$ to $m'$, there is the same type of walk from $m'$ to $l$  with $m'\in M\cup M_1$. In addition, $l\in M_1$ remains the same as an endpoint of subsections on which there are nodes on $\pi_3$ that are not in $M\cup M_1$.

{\bf In $\alpha_{CMG}(G;M\cup M_1,\varnothing)$:} By Lemma \ref{lem:10}, all subpaths of $\pi_3$ of form $\pi'$ are replaced by the $k'l$ arrows or arcs respectively. Therefore, there is an endpoint-identical walk whose inner sections are all non-collider and whose inner nodes are all in $M\cup M_1$. By Lemma \ref{lem:1}, we conclude that there is an endpoint-identical (i.e.\ the same type of) edge between $i$ and $j$.

{\bf ($\Leftarrow$)} Suppose that there is an edge between $i$ and $j$ in $\alpha_{CMG}(G;M\cup M_1,\varnothing)$. Starting from this edge, we discuss the type of path or walk that exists between $i$ and $j$ in every graph generated by different steps of Algorithm \ref{alg:2}:

{\bf In  the graph generated before applying step 2 of Algorithm \ref{alg:2} to $G$ for $M\cup M_1$:} By Lemma \ref{lem:1}, there exists an endpoint-identical walk
$\pi$ between $i$ and $j$  whose inner sections are all non-collider and inner nodes are all in $M\cup M_1$.

{\bf In $G$:} By Lemma \ref{lem:11}, there is a new walk, denoted by $\pi_1$.  Define $l$ also as defined in the lemma, and notice that in this case $l$ is both in $M\cup M_1$ and a child of $m_1\in M\cup M_1$.

{\bf In the graph generated after applying step 1 of Algorithm \ref{alg:2} to $G$ for $M$:} All subpaths of $\pi_1$ of the mentioned form and properties $\pi'$  where $l$ is a child of $M$ can be replaced by $kl$ arrows or lines respectively.

{\bf In $\alpha_{CMG}(G;M,\varnothing)$:} Now the generated walk can be partitioned into subwalks with endpoints in outside $M$ and all inner nodes in $M$ (there might be single edges in the partition). All these subwalks with lengths more than two satisfy the conditions of Lemma \ref{lem:1} for $M$. Hence, there exist endpoint-identical edges between the endpoints of the subwalks. These edges form a walk, which is denoted by $\pi_2$.

{\bf In the graph generated after applying step 1 of Algorithm \ref{alg:2} to $\alpha_{CMG}(G;M,\varnothing)$ for $M_1$:} Since there are no collider sections on $\pi_1$, and because of endpoint-identicality, there are no collider sections on $\pi_2$. In addition, the endpoints $l$ (as defined)  of subpaths of $\pi_2$ whose members may not be in $M_1$, are  children of $M_1$. Therefore, again by applying step 1 of the algorithm for $M_1$ we obtain a walk with all inner nodes in $M_1$.

{\bf In $\alpha_{CMG}(\alpha_{CMG}(G;M,\varnothing);M_1,\varnothing)$:} Now by applying Lemma \ref{lem:1} to the generated walk, we obtain an endpoint-identical (and hence the same type $ij$ edge as the original $ij$ edge).
\end{proof}
\begin{proof}[Proof of Proposition \ref{prop:2vn}]
{\bf First, we prove that every CG $G$ is mapped into $\mathcal{H}$:} By proposition \ref{prop:2}, we know that the generated graphs are CMGs. We consider each case separately:

\emph{Suppose that there is a collider trislide of form $k\arc i\ful\dots\ful j\fla l$ in the generated graph $\alpha_{CMG}(G;M,\varnothing)$.} We go through how this trislide has been generated by steps of Algorithm \ref{alg:2}.

\emph{In the graph generated before applying step 2 of Algorithm \ref{alg:2}:} Since by step 2 of Algorithm \ref{alg:2} only case 7 of Table \ref{tab:1} can generate lines, by an inductive argument it is clear that between $i$ and $j$ there is a section. By Lemma \ref{lem:1}, instead of the arrow from $l$ to $j$, there is a walk with non-collider sections and inner nodes in $M$ such that there is an arrowhead at the endpoint section containing $j$ (say from node $r$, which may be $l$).

In addition, notice that $G$ is a CG and by step 1 of Algorithm \ref{alg:2}, no arc is generated from trislides that do not contain arcs. This fact together with Lemma \ref{lem:1} implies that there is a  walk between $i$ and $k$ that only contains lines and arrows, and, on this walk, there is an arrowhead at the endpoint section containing $i$ (say at node $o$, which may be $i$ and has a parent in $M$).

By considering the path between $r$ and $o$, we conclude that by step 1 of  Algorithm \ref{alg:2} (case 8 of Table \ref{tab:1}), an arrow from $r$ to $o$ is generated.

\emph{In $\alpha_{CMG}(G;M,\varnothing)$:} Now by Lemma
\ref{lem:1}, and considering the walk with  non-collider sections and inner node in $M$ that connects $l$, $r$, $o$, and $i$, an arrow from $l$ to $i$ is generated.

 \emph{Suppose that there is a collider trislide of form $k\arc i\ful\dots\ful j\arc l$ in the generated graph:}  $r$ and $o$ can be defined in the same way as in the previous case. Notice that in this case on the walk (obtained by Lemma \ref{lem:1}) there is an arrowhead at the section containing $l$. By a similar argument to that in the previous case, we conclude that there is an arc generated between $l$ and $i$ in the generated graph. By the symmetry in the path, one can similarly obtain  an arc between $k$ and $j$. Furthermore,  by Lemma \ref{lem:1}, and considering the walk with  non-collider sections and inner node in $M$ that connects $j$, $r$, $o$, and $i$,  there exists an arc between $i$ and $j$ in the generated graph, since,  on this walk, there are arrowheads at both sections that contain $i$ an $j$.

{\bf Now we prove that the function is surjective:} Consider an arbitrary chain mixed graph $H$ in $\mathcal{H}$. Define a chain graph $G$ from $H$ as follows: keep all arrows and lines of $H$ in $G$ and replace arcs $ij$ with $i\fla m \fra j$; and define a subset $M$  of the node set of $G$ as the set of all such $m$.

\emph{We first prove that $G$ is a CG:} It only contains the two desired types of edges. In addition, it does not contain semi-directed cycles that contains an arrow since if, for contradiction, it does then it must contain the tripath $i\fla m \fra j$, which is impossible.

\emph{We now prove that $\alpha_{CMG}(G;M,\varnothing)=H$:} The changes that might occur by step 1 of Algorithm \ref{alg:2} are only when, in $H$, there are the two types of collider trislides in properties 1 and 2, which correspond to the walks $k\fla m \fra i\ful\dots\ful j\fla l$ and $k\fla m_1\fra i\ful\dots\ful j\fla m_2 \fra l$ in $G$. In the former case, the generated arrow from $l$ to $i$ exists in $H$. In the latter case, an arrow from $m_2$ to $i$ is generated, but since $m_2$ is only adjacent to $j$ and $l$, in the next step of the algorithm, it can only generate $il$ and $ij$ arcs, both of which exist in $H$; the same argument also works for the generated arrow from $m_1$ to $j$. In addition, step 9 is not applied since there are no arcs in $G$. By step 2 of the algorithm, the only type of tripath with inner node in $M$ is case 4 of Table \ref{tab:1} (except those that are already discussed). These tripaths obviously turn into the arcs existing in $H$, and no other edge is generated.
\end{proof}
\begin{proof}[Proof of Theorem \ref{thm:2}]
We need to prove that $A\dse_cB\cd C_1$ in $G\iff A\dse_cB\cd C_1$ in $\alpha_{CMG}(G;M,\varnothing)$.

{\bf ($\Rightarrow$)} Suppose that there is a $c$-connecting walk $\pi$ given $C_1$ between $i$ and $j$ in $G$. Consider all maximal subwalks of $\pi$ whose inner sections are all non-collider, endpoints are not in
$M$, and inner nodes are all in $M$.  Notice that all nodes of
$\pi$ that are in $M$ are included in these subwalks since no collider section on $\pi_1$ has all nodes in $M$. Denote such a subwalk by $\varpi$.

{\bf In the generated graph after applying step 1 of Algorithm \ref{alg:2}:} First consider the case where the endpoints of $\varpi$ are the same node $l$. Sections on $\varpi$ are non-collider, and hence, the edge between $l$ and an endpoint of $\varpi$ (call it $m$) is an arrow from $m$ to $l$. We can easily obtain a shorter $c$-connecting walk by removing $\varpi$ from $\pi$  if, by doing so, $l$ is on a collider section or on a non-collider section with no node in $C_1$.  If that is not the case then there exists $l\fla m\fra l\ful\cdots\ful \circ\fla\, k$ or $l\fla m\fra l\ful\cdots\ful \circ\arc\, k$, where $l\not\in C_1$ but an inner node of the section containing $l$ is in $C_1$. (Notice that if $l$ is $i$ or $j$ then one can easily remove $m$ from the walk.) By step 1, there is a generated $lk$ edge. We replace all these walks with the generated edge and call the resulting walk $\pi_1$. Because the generated edges are endpoint-identical to the subwalks, $\pi_1$ is $c$-connecting.

{\bf In the generated graph after applying step 2 of Algorithm \ref{alg:2}:}  The subwalks of $\pi_1$ with the property mentioned above now have distinct endpoints. By Lemma \ref{lem:1}, instead of these subwalks, there are endpoint-identical edges in $\alpha_{CMG}(G;M,\varnothing)$. By replacing all the subpaths with these edges, we obtain a walk $\pi_2$. Walk $\pi_2$ exists in $\alpha_{CMG}(G;M,\varnothing)$ since there are no members of $M$ on $\pi_2$. In addition, $\pi_2$ is $c$-connecting given $C_1$ since, because of endpoint-identicality of the generated edges to the subwalks, every node
that is an inner node of a collider or a non-collider section on $\pi_2$ is an inner
node of a collider or a non-collider section on $\pi_1$, and no node in $C_1$ on $\pi_1$ has been taken out.
%

{\bf ($\Leftarrow$)} Suppose that there is a $c$-connecting walk $\pi$ given $C_1$ between $i$ and $j$ in $\alpha_{CMG}(G;M,\varnothing)$. We show what types of walks generated $\pi$ at each step of Algorithm \ref{alg:2}.

{\bf In the graph before applying step 2 of Algorithm \ref{alg:2}:} By Lemma \ref{lem:1}, for every edge $kl$ on $\pi$, there is an endpoint-identical walk $\pi'$ between $k$ and $l$  with the stated properties in the lemma. By replacing every edge on $\pi$  by such $\pi'$, we obtain a walk $\pi_1$. We prove that $\pi_1$ is $c$-connecting given $C_1$: Notice that $\pi'$ is obviously $c$-connecting. In addition, because of endpoint-identicality, for a replaced edge $kl$, if $l$ is an inner node of a collider or a non-collider section, after the replacement, it remains an inner node of a collider or non-collider section respectively, and all added nodes are in $M$.

{\bf In $G$, before applying step 1 of Algorithm \ref{alg:2}:}
Now a $uv$ edge on $\pi_1$ might have been replaced by a path by step 1 of the algorithm, where $u$ is a child of $m\in M$. By all such replacements, we obtain a larger walk $\pi_2$. Again, because of endpoint-identicality, if $u$ is on a collider section or a non-collider section $\rho_1$ on $\pi_1$ then it remains on a (possibly larger) collider section or a non-collider section $\rho_2$ on $\pi_2$ respectively. If $\rho_2$ is non-collider and  all inner nodes of the new path are outside $C_1$ then it is clearly open on $\pi_2$. If $\rho_2$ is non-collider with a node in $C_1$ then we modify $\pi_2$ by adding the subwalk $\langle u,m,u\rangle$ (i.e.\ the arrow from $m$ to $u$ in both directions) to $\pi_2$. Now the subpath of $\rho_2$ between $v$ and $u$ becomes a collider section and open on $\pi_2$, and the rest of $\rho_2$ (with an arrow pointing to it from $m$) remains a non-collider section and open. If $\rho_2$ is a collider, it is clearly open since there is already a node in $C_1$ on $\rho_1$. Therefore, by an inductive argument, $\pi_2$ is a $c$-connecting walk.
\end{proof}
\begin{proof}[Proof of Lemma \ref{lem:2}]
{\bf ($\Leftarrow$)} Suppose that there exists a walk $\pi$ between $i$ and $j$ in the generated graph after step 2  whose inner sections are all collider and in $C\cup\ant(C)$, and endpoint sections contain a single node. We prove the result by induction on the number of edges of $\pi$. If it is $1$ then we are clearly done. If it is $n>1$ then consider the trislide $\tau=\langle i\rho k\rangle$ on $\pi$, where $\rho$ is a section.  By step 3 of the algorithm, an endpoint-identical edge $ik$ is generated. Notice that $ik$ is either an arrow or an arc unless possibly $k=j$. Now by replacing $\tau$ by the $ik$ edge, we obtain a shorter walk with the same properties.  By the induction hypothesis, we obtain the result.

{\bf ($\Rightarrow$)} Suppose that there is an edge between $i$ and $j$ in the graph generated after step 3 of Algorithm \ref{alg:3}. If this edge were  generated by step 3 of Algorithm \ref{alg:3} then it would be generated by one of the first three trislides in Table \ref{tab:2} in an iteration of step 3 of the algorithm. Each arrow or arc  on the trislide may now have been generated by a trislide with inner nodes in $C\cup\ant(C)$ (since no generated line can be used in the iterations). Since the trislides are endpoint-identical to the generated edge, it is implied that all sections remain collider. By an inductive argument, we imply that, in the graph generated after applying step 2 of the algorithm, there is an endpoint-identical walk between $i$ and $j$ whose inner nodes are in $C\cup \ant(C)$ and all sections are collider. In addition, $i$ and $j$ are clearly not adjacent to a line on this walk, i.e., endpoint sections contain a single node.
\end{proof}
\begin{proof}[Proof of Lemma \ref{lem:20}]
The first result for step 4 is trivial, and for step 3 follows directly from Lemma \ref{lem:2}. This implies that if a generated line lies on a collider section after step 3 then since $j\in S$, by step 4, all arrowheads at the section will be removed.
\end{proof}
\begin{proof}[Proof of Lemma \ref{lem:21}]
One direction of the proof is obvious since steps 1, 2, and 3 of Algorithm \ref{alg:3} do not remove or replace any edges, and by removing an arrowhead at an arrow pointing to  $i$ by step 4, no new node can become an anterior of $i$. Thus, suppose that $i\in\ant(C)$  after step 4 of the algorithm. We go back on the steps of the algorithm in order to show that $i$ has been in $\ant(C)$.

{\bf Before applying step 4 of Algorithm \ref{alg:3}:} Suppose that there is a node $k$ on  the semi-directed path $\pi$ from $i$ to $C$ such that, on $\pi$, there is an arrowhead at $k$ on the opposite direction of $\pi$. In addition, suppose that this arrowhead has been removed by step 4. It then holds that $k\in C\cup\ant(C)$. By considering the closest of such nodes to $i$ on $\pi$, $i$ is an anterior of $k$, and consequently $C$.

{\bf Before applying step 3 of Algorithm \ref{alg:3}:} Consider the closest arrow to $i$  on $\pi$ that is generated by step 3. The result then follows from Lemma \ref{lem:20}.

{\bf Before applying step 2 of Algorithm \ref{alg:3}:} The only possible arrow on $\pi$ (say from $k$ to $l$) can be generated by step 2 (case 4 of Table \ref{tab:2}). This implies that $k\in\ant(l)$. By an inductive argument, this implies the result.
\end{proof}
\begin{proof}[Proof of Proposition \ref{prop:3}]
Graphs generated by Algorithm \ref{alg:3} have the three desired types of edges. We prove that there is no semi-directed cycle with an arrow in a generated chain mixed graph from $G$.  Suppose, for contradiction, that a generated graph does contain a semi-directed cycle $\pi$ with an arrow. Since $\pi$ does not exist in $G$, at least one arrow, say from $j$ to $i$, or a line, say between $k$ and $l$ has been generated by Algorithm \ref{alg:3}. If $ij$ or $lk$ has been generated by steps 3 or 4 of the algorithm then by Lemma \ref{lem:20}, $j,k,l\in S$ in $G$. This implies that there should be no arrow on $\pi$, a contradiction.

Thus, the only option that is left is that $ij$ has been generated by step 2, case 4 of Table \ref{tab:2}. In this case $j\in\ant(i)$ in $G$ with an arrow existing on the directed path from $j$ to $i$. By considering all arrows generated by this step of the algorithm on $\pi$, we conclude that there is a semi-directed cycle with an arrow in $G$, a contradiction.
\end{proof}
\begin{proof}[Proof of Lemma \ref{lem:2n}]
{\bf We first prove that there is an $ij$ edge in $\alpha_{CMG}(G;\varnothing,C)$ if and only if there is a walk as described in the lemma in $G$:}

{\bf ($\Rightarrow$)} Suppose that in $\alpha_{CMG}(G;\varnothing,C)$ there is an edge between $i$ and $j$. We will follow how this edge might have been generated by the steps of Algorithm \ref{alg:3}.

{\bf In  the graph generated before applying step 4:} It is clear that there is an $ij$ edge.

{\bf In  the graph generated before applying step 3:} Now, by Lemma \ref{lem:2}, there
exists an endpoint-identical walk $\pi$ between $i$ and $j$ to the edge $ij$  whose inner sections are all collider and  in $S$, and whose endpoint sections contain a single node. Notice that this means that all edges on $\pi$ are either lines or arcs except possibly those containing $i$ and $j$.

{\bf In $G$:} By replacing arcs or arrows on $\pi$ by endpoint-identical paths (provided in cases 4 and 5 of Table \ref{tab:2}), only collider sections on $\pi$ become larger. The newly added nodes to the sections will obviously be in $S$ since they are anteriors of the rest of the section, which is in $S$ -- the only exception is when there is an arrowhead at $i$ (or $j$) and the section containing $i$ gets larger. In this case, $i\in\spo(k)$, for $k\in S$.

{\bf ($\Leftarrow$)} Suppose that in $G$, there exists a walk $\pi$ as described in the lemma. The edges of $\pi$ are all arcs and lines except possibly those including $i$ and $j$. We will go through how this walk changes by the steps of Algorithm \ref{alg:3}.

{\bf In  the graph generated after applying step 2:} The endpoint sections turn into single nodes, and other sections may get shortened, but since the generated edges are endpoint-identical to the generating paths (provided in cases 4 and 5 of Table \ref{tab:2}), inner sections of the resulting walk are still collider. Lemma \ref{lem:21} implies that the inner sections stay in $S$.

{\bf In $\alpha_{CMG}(G;\varnothing,C)$:} The generated walk in the previous step satisfies the conditions of Lemma \ref{lem:2}. Hence, there is an $ij$ edge generated by step 3, which keeps existing after step 4.

{\bf  We now prove the second claim in the lemma:} Since all generated edges by steps 2 and 3 of the algorithm (all cases of Table \ref{tab:2}) are endpoint identical to their generating paths, the generated edge after step 3 and the walk in $G$ are endpoint-identical. Step 4 changes endpoint-identicality only when it removes the arrowhead at $i$, which always and only happens when $i\in\ant(C)$.
%
\end{proof}
\begin{proof}[Proof of Theorem \ref{thm:3n}]
Notice that $i,j\notin C\cup C_1$. {\bf We first prove that there is an $ij$ edge in $\alpha_{CMG}(\alpha_{CMG}(G;\varnothing,C);\varnothing,C_1)$ if and only if there is an $ij$ edge in $\alpha_{CMG}(G;\varnothing, C\cup C_1)$:}

By Lemma \ref{lem:2n}, there is an edge between $i$ and $j$ in $\alpha_{CMG}(\alpha_{CMG}(G;\varnothing,C);\varnothing,C_1)$ if and only if there is a walk $\pi$ as described in the lemma between $i$ and $j$ in $\alpha_{CMG}(G;\varnothing,C)$ with inner sections in $S_1=C_1\cup\ant(C_1)$.

Notice that by Lemma \ref{lem:20}, lines on the inner sections of $\pi$ exist in $G$. In addition, there at most two arrows might exist on $\pi$, which are from the endpoints $i$ and $j$. Now again by Lemma \ref{lem:2n}, instead of a $kl$ arc on $\pi$, in $G$, there is an endpoint-identical walk $\pi'$ as described in the lemma between $k$ and $l$ with inner sections in $S=C\cup\ant(C)$. By replacing $kl$ by $\pi'$, one obtains a walk with the same properties as in Lemma \ref{lem:2n} for $S\cup S_1$. Inductively, we replace all such $kl$ arcs. We also replace a possible arrow (say from $i$ to $h$) by a walk with properties as described in Lemma \ref{lem:2n}, where there might be an arrowhead at $i$ with $i\in\ant(C)$. By all these replacements, one obtains a walk $\pi_1$ in $G$. Since conditions of Lemma \ref{lem:2n} are both necessary and sufficient, it holds that there is the walk $\pi$ in $\alpha_{CMG}(G;\varnothing,C)$ if and only if there is the walk $\pi_1$ in $G$.

Walk $\pi_1$ satisfies the properties in Lemma \ref{lem:2n} for $S\cup S_1$. Again by Lemma \ref{lem:2n},  there is the walk $\pi_1$ in $G$ if and only if there is an $ij$ edge in $\alpha_{CMG}(G;\varnothing, C\cup C_1)$.

{\bf  We now prove that the $ij$ edge is the same in both graphs:}  We only need to show that there is an arrowhead at $i$ on the $ij$ edge in $\alpha_{CMG}(\alpha_{CMG}(G;\varnothing,C);\varnothing,C_1)$ if and only if there is an arrowhead at $i$ on the $ij$ edge in $\alpha_{CMG}(G;\varnothing, C\cup C_1)$. This follows from the second part of Lemma \ref{lem:2n} and the fact that if $i\in\ant(C)\cup\ant(C_1)$ in $G$ then there is no arrowhead at the $ij$ edge in $\alpha_{CMG}(\alpha_{CMG}(G;\varnothing,C);\varnothing,C_1)$ or $\alpha_{CMG}(G;\varnothing, C\cup C_1)$. Below we prove the latter claim:

The result for  $\alpha_{CMG}(G;\varnothing, C\cup C_1)$ is again clear by Lemma \ref{lem:2n}. We now condider $\alpha_{CMG}(\alpha_{CMG}(G;\varnothing,C);\varnothing,C_1)$. If $i\in\ant(C)$ then there is no arrowhead at $i$ on $\pi$ in $\alpha_{CMG}(G;\varnothing,C)$. If $i\in\ant(C_1)\setminus\ant(C)$ then consider the semi-directed path from $i$ to a member of $C_1$ in $G$. This path remains intact in  $\alpha_{CMG}(G;\varnothing,C)$ since $i\notin\ant(C)$. Hence, the arrowhead at $i$ on the $ij$ edge will be removed in $\alpha_{CMG}(\alpha_{CMG}(G;\varnothing,C);\varnothing,C_1)$.

\end{proof}
\begin{proof}[Proof of Theorem \ref{thm:3}]
We prove that $A\dse_cB\cd C\cup C_1$ in $G$ if and only if $A\dse_cB\cd C_1$ in $\alpha_{CMG}(G;\varnothing,C)$.

{\bf ($\Leftarrow$)} Suppose that there is a $c$-connecting walk $\pi$ given $C\cup C_1$ between $i$ and $j$ in $G$. We apply the steps of Algorithm \ref{alg:3} to this walk. Consider all maximal subwalks of $\pi$ whose inner sections are all collider and in  $C$, and endpoints are single nodes and not in $C$.  Notice that all nodes of $\pi$ that are in $C$ are included in these subwalks since no non-collider section on $\pi$ has a node in $C$. Denote such a subwalk by $\varpi$.

{\bf After applying step 2:} First consider the case where the endpoints of $\varpi$ are the same node $l$. Sections on $\varpi$ are collider, and hence, the edge between $l$ and an endpoint of $\varpi$ (call it $c$) has an arrowhead at $c$. We can easily obtain a shorter $c$-connecting walk by removing $\varpi$ from $\pi$  if, by doing so, $l$ is on a collider section or on a non-collider section with no node in $C\cup C_1$. First, this implies that the $cl$ edge is an arc. In addition, if that is not the case then there exists $l\arc c\arc l\ful\cdots\ful \circ\fla\, k$ or $l\arc c\arc l\ful\cdots\ful \circ\arc\, k$, where $l\not\in C\cup C_1$ but an inner node of the section containing $l$ is in $C_1$. (Notice that if $l$ is $i$ or $j$ then one can easily remove $c$ from the walk.) By step 2, there is a generated $lk$ edge. We replace all these walks with the generated edge and call the resulting walk $\pi_1$. Because the generated edges are endpoint-identical to the subwalks, $\pi_1$ is $c$-connecting.

{\bf After applying step 3:}  By Lemma \ref{lem:0}, there is an alternative $c$-connecting walk $\pi'_1$ to $\pi_1$, where all sections are paths and inner nodes of collider sections are in $C\cup C_1\cup(\ant(C)\cup\ant(C_1))$.  Consider all maximal subwalks of $\pi'_1$ whose inner sections are all collider and in  $C\cup\ant(C)$, and endpoints are single nodes and not in $C$.  Because of the previous step, the endpoints of such subwalks are distinct nodes. Now, by Lemma \ref{lem:2}, instead of these subwalks, there are endpoint-identical edges. By replacing all the subwalks with these edges, we obtain a walk $\pi_2$. Walk $\pi_2$ is $c$-connecting given $C_1$ since generated edges on $\pi_2$ are endpoint-identical to the subpaths on $\pi'_1$ that have been replaced.

{\bf After applying step 4:} By this step, no collider sections turn into a non-collider one on $\pi_2$ since if an arrowhead on a node $k$ is removed then $k\in\ant(C)$ in $G$ and so are all inner nodes of the section that contains $k$. Hence, $k$ cannot be on $\pi_2$ by how $\pi_2$ is generated. Therefore, $\pi_2$ is a $c$-connecting walk given $C_1$ in $\alpha_{CMG}(G;\varnothing,C)$.

{\bf ($\Rightarrow$)} Suppose that there is a $c$-connecting walk $\pi$ given $C_1$ between $i$ and $j$ in $\alpha_{CMG}(G;\varnothing,C)$. In $G$, we obtain a walk $\pi'$ by replacing every edge on $\pi$ with the corresponding walks described in Lemma \ref{lem:2n}. All these generated walks by edges of $\pi$ are $c$-connecting given $C\cup C_1$ themselves. Hence, if their endpoints  are open then $\pi'$ would be $c$-connecting given $C\cup C_1$.

If a generated subwalk on $\pi'$ is endpoint-identical to the generating edge on $\pi$ with endpoint sections containing a single node then it is open. Hence, we need to consider two cases where this does not happen for a generated subwalk:

1) When the endpoint sections of the generated walks contain more than a node, we know that there is an arrowhead at the section, and the endpoint $k$ is a spouse of $s\in S$. It is possible that the endpoint section $\rho$ is not open on $\pi'$ (but the corresponding edge is open on $\pi$) if it is a non-collider with a node in $C$. In this case add $\langle k,s,k\rangle$ (i.e., repeating the $ks$ edge twice) instead of $k$ to $\pi'$. This makes $\rho$ collider and also adds a collider section $s$ (containing a single node) and one non-collider section containing $k$, which  are all open.

2) We know that the generated walks on $\pi'$ and the generating edges on $\pi$ are endpoint-identical except when there is an arrowhead at the endpoint section $\rho'$ containing $l$ and there is a semi-directed path $\varpi$ from $l$ to $c\in C$ in $G$. In this case, add $\langle \varpi,\varpi^r\rangle$ instead of $l$ to $\pi'$ (i.e.\ go from $l$ to $c$ and come back to $l$ on $\varpi$). By this method, we split the collider $\rho'$ at $l$ into two subpaths, both of which are non-colliders, and obtain other open non-collider sections along $\varpi$ and a collider section $c$.
\end{proof}
\begin{proof}[Proof of Proposition \ref{prop:1}]
The generated graphs obviously contain only lines and arrows, thus it is enough to prove that they do not contain semi-directed cycles with an arrow. Suppose, for contradiction, that a generated graph does
contain a semi-directed cycle $\pi$ with an arrow. If a line $ij$ on $\pi$ has been generated by step 4 then $i,j\in S$ in $G$ and, therefore, all nodes on $\pi$ are in $S$ . This implies that there is no arrow on $\pi$, a contradiction. If a line $kl$ has been generated by step 3 then it is easy to see that both $k,l\in S$, and again  there is no arrow on $\pi$, a contradiction. Therefore, all lines on $\pi$ exist in the original graph, and no arrows are generated by the algorithm. Hence, $\pi$ exists in the original graph, a contradiction.
\end{proof}
\begin{proof}[Proof of Lemma \ref{lem:vvn}]
We show that for any choice of $C$, $i\dse j\cd C$ dos not hold:  Suppose that there is an arrow from an inner node $k$ to $j$. If any of the inner nodes is in $C$ then $i$ and $j$ are dependent given $C$. If no inner node is in $C$ then the subwalk between $i$ and $k$ in addition to the $kj$ arrow constitutes a connecting walk given $C$.
\end{proof}
\begin{proof}[Proof of Lemma \ref{lem:vu}]
{\bf We prove the first claim:}

{\bf ($\Rightarrow$)} Suppose that in $\alpha_{CMG}(\alpha_{CMG}(G;\varnothing,C);M,\varnothing)$ there is an edge between $i$ and $j$.

{\bf In the graph generated before applying step 2 of Algorithm \ref{alg:2} to $\alpha_{CMG}(G;\varnothing,C)$:} By lemma \ref{lem:1}, there exists a walk $\pi$ between $i$ and $j$  whose inner sections are all non-collider and inner nodes  are all in $M$.

{\bf In $\alpha_{CMG}(G;\varnothing,C)$:} By Lemma \ref{lem:11}, there is a walk $\pi_0$ between $i$ and $j$ with the same non-collider sections. In addition, every node on $\pi_0$ on section $\rho$ that is not in $M$ is on a subsection with an endpoint that is the endpoint of $\rho$ as well with an arrowhead pointing to it from the other adjacent node on $\pi_0$. The other endpoint $h$ is in $M$ and a child of a member of $M$.

{\bf In $G$:} For every edge $kl$ on $\pi_0$, by Lemma \ref{lem:2n}, there exists a walk $\pi'$ between $k$ and $l$  whose inner sections are all collider and in $C\cup\ant(C)$. We denote the walk in this graph that consists of all such adjacent $\pi'$ of $\pi_0$ by $\pi_1$. Even if the endpoint sections of $\pi'$ are not single elements or $\pi'$ is not endpoint-identical to the $kl$ edge, all the existing non-collider sections remain non-collider (although some sections might become larger). It is then observed that all non-collider sections on $\pi_1$ have all inner nodes outside $C$, and all collider sections have inner nodes in $C\cup\ant(C)$. In addition, every node on $\pi_1$ on section $\rho'$ that is not in $M$ is on a subsection with an endpoint that is the endpoint of $\rho'$ as well with an arrowhead pointing to it from the other adjacent node on $\pi_1$. The other endpoint $h$ is in $M$ and either a child of a member of $M$ or a spouse of a member of $C\cup\ant(C)$.

{\bf ($\Leftarrow$)} Suppose that there is a walk between $i$ and $j$ in $G$ with the two mentioned properties. In place of this walk, we have the following walks in the following graphs:

{\bf After applying step 1 of Algorithm \ref{alg:2} to $G$:} By this step it can bee seen that all subwalks containing non-collider sections outside $M$ with an endpoint that is a child of $M$ get closed, and, therefore, we obtain a walk on which (i) all nodes on collider sections are in $C\cup\ant(C)$; (ii) (a) all nodes on non-collider sections are in $M$ or (b) on the non-collider section one endpoint is in $M$ and a spouse of a node in $C\cup\ant(C)$, and the other endpoint has an arrowhead at it from the adjacent node on the walk.

{\bf In $\alpha_{CMG}(G;M,\varnothing)$:} By Lemma \ref{lem:1}, we obtain a walk on which all sections are collider and in $C\cup\ant(C)\cup\ant(j)$. Notice that the spouses of the endpoints of non-collider sections in the previous walk, which are in $C\cup\ant(C)$, appear on the generated walk.

{\bf In $\alpha_{CMG}(G;M,C)$:} By Lemma \ref{lem:2n}, we obtain an edge.
%

{\bf We now prove the second claim:} We go through the corresponding walks in the intermediate graph, provided above. By lemma \ref{lem:2n}, the $ij$ edge in $\alpha_{CMG}(G;M,C)$ and the corresponding walk in $\alpha_{CMG}(G;M,\varnothing)$ remain endpoint-identical except when there is an arrowhead at the endpoint section containing, say, $i$, and $i\in \ant(C)$ in $\alpha_{CMG}(G;M,\varnothing)$. This walk, by Lemma \ref{lem:1}, is endpoint-identical to the corresponding walk in the graph generated after applying step 1 of Algorithm \ref{alg:2} to $G$. Since the anterior set does not change at this step and the next step in $G$, and since step 1 of Algorithm \ref{alg:2} generates endpoint-identical edges, the result follows for the corresponding walk in $G$.
\end{proof}
\begin{lemma}\label{lem:an1}
For a chain mixed graph $G$ and $M$ and $C$ subsets of its node set, if $i\in\ant(j)$ in $\alpha_{CMG}(\alpha_{CMG}(G;\varnothing,C);M,\varnothing)$ then $i\in\ant(C\cup \{j\})$ in $G$.
\end{lemma}
\begin{proof}
The proof follows from Lemma \ref{lem:vu} by the following observations: A line between $k$ and $l$ or an arrow from $k$ to $l$ on the semi-directed walk from $i$ to $j$ in $\alpha_{CMG}(\alpha_{CMG}(G;\varnothing,C);M,\varnothing)$ is not endpoint-identical to the corresponding walk $\pi$ in $G$ if and only if $k\in\ant(C)$ in $G$. If they are endpoint-identical then start from $k$ and move towards $l$ on $\pi$. At each step we either reach a collider section and conclude that $k\in\ant(C)$, or we finally reach $l$ and conclude that $k\in\ant(l)$. By an inductive argument on the nodes of $\pi$, we obtain the result.
\end{proof}
\begin{proof}[Proof of Proposition \ref{prop:l}]
{\bf We first prove that there is an $ij$ edge in $\alpha_{CMG}(\alpha_{CMG}(G;M,\varnothing);\varnothing,C)$ if and only if there is an $ij$ edge in $\alpha_{CMG}(\alpha_{CMG}(G;\varnothing,C);M,\varnothing)$:} We go through Algorithms \ref{alg:2} and \ref{alg:3} to follow the types of walks corresponding to the $ij$ edge in any of these graphs in each step of the algorithms.

{\bf ($\Rightarrow$)} Suppose that  in $\alpha_{CMG}(\alpha_{CMG}(G;M,\varnothing);\varnothing,C)$ there is an edge  between $i$ and $j$.


{\bf In $\alpha_{CMG}(G;M,\varnothing)$:} By Lemma \ref{lem:2n}, there is a walk $\pi$ between $i$ an $j$ with the properties described in the lemma.

{\bf In the graph generated before applying step 2 of Algorithms \ref{alg:2} to $G$:} For every edge $kl$ on $\pi$, by Lemma \ref{lem:1}, there exists an endpoint-identical walk $\pi'$ between $k$ and $l$ whose inner sections are all non-collider and inner nodes are all in $M$. We denote the walk that consists of all such adjacent $\pi'$ by $\pi_0$. It is easy to observe that all collider sections are in $C\cup\ant(C)$. In addition, either the endpoint sections of $\pi_0$ still satisfy the conditions of Lemma \ref{lem:2n}, or the endpoints that are not single elements become children of members of $M$.

{\bf In $G$:} By Lemma \ref{lem:11}, there exists another walk $\pi_1$, on which,  all collider sections are in $C\cup\ant(C)$. In addition, collider and non-collider sections remain intact. In addition, it can be seen that on $\pi_1$, the conditions for endpoint sections described in the previous paragraph still hold.

%

{\bf In $\alpha_{CMG}(\alpha_{CMG}(G;\varnothing,C);M,\varnothing)$:} The walk described in the previous paragraph in $G$ satisfies the conditions of Lemma \ref{lem:vu}. Hence, by this lemma, we obtain the result.


{\bf ($\Leftarrow$)} Suppose that in $\alpha_{CMG}(\alpha_{CMG}(G;\varnothing,C);M,\varnothing)$ there is an edge between $i$ and $j$. By Lemma \ref{lem:vu}, there is a walk $\pi_1$ as described in the lemma in $G$. We now continue to check how this walk alters along the steps of the relevant algorithms:

{\bf In the graph generated after applying step 1 of Algorithm \ref{alg:2} to $G$:} All maximal subsections of non-collider sections whose nodes are outside $M$, but an endpoint $l$ is in $M$ and a child of $M$ can be replaced by an endpoint-identical edge. By all such replacements, we obtain a walk $\pi_2$, which contains collider sections in $C\cup\ant(C)$ and non-collider sections outside $C$. In addition,  every node on $\pi_2$ on section $\rho$ that is not in $M$ is on a subsection with an endpoint that is the endpoint of $\rho$ as well with an arrowhead pointing to it from the other adjacent node on $\pi_2$. The other endpoint $h$ is in $M$ and  a spouse of a member of $C\cup\ant(C)$.

{\bf In $\alpha_{CMG}(G;M,\varnothing)$:}  First consider a non-collider trislide $\langle r,\rho', q\rangle$ where $\rho'$ has members outside $M$. In addition, say $r$ is the endpoint of $\rho'$ with an arrowhead pointing to it from the other adjacent node on $\pi_2$. Consider the node $h$ as defined in the above paragraph, which is a spouse of $s\in C\cup\ant(C)$. Denote the adjacent node to $h$ closer to $r$ by $t$ and the adjacent node to $h$ closer to $q$ by $v$.  By this step, an edge between $t$ and $v$ as well as $ts$ and $sv$ arcs are generated.

In addition, by using Lemma \ref{lem:1}, we replace the maximal subwalks of $\pi_2$ that contain only non-collider sections and in which all nodes are in $M$, but endpoints are outside $M$, by the generated endpoint-identical edges. By all these replacements, we obtain a walk $\pi_3$  that contains collider sections with nodes in $C\cup\ant(C)$ and non-collider sections outside $C$. In particular, we obtain an $sq$ arc as well as an arrow from $t$ to $q$.

{\bf In $\alpha_{CMG}(\alpha_{CMG}(G;M,\varnothing);\varnothing,C)$:} By  Lemma \ref{lem:2n}, instead of all subwalks of $\pi_3$ that contain inner collider sections, there exists an edge. In addition, for non-collider sections, the collider tripath $\langle t,s,q\rangle$ (described in the above paragraph) generates a $tq$ arc. Because of the arrow from $t$ to $q$ and the subwalk of the trislide between $r$ and $t$, and by Lemma \ref{lem:vvn}, we conclude that  the graph is not maximal except when there is an endpoint-identical edge between $r$ and $q$. Therefore, by an inductive argument, there is an edge between the endpoints of $\pi_3$.

{\bf We now prove that the $ij$ edge is of the same type in both graphs:}  For every  graph generated by a step of the algorithm, we discussed a walk between $i$ and $j$ in both directions of the proof above. We focus on the arrowhead pointing to $i$ on these walks:


By Lemma \ref{lem:2n}, there is no arrowhead pointing to $i$ on the $ij$ edge in $\alpha_{CMG}(\alpha_{CMG}(G;M,\varnothing);\varnothing,C)$ if and only if there is no arrowhead pointing to $i$ or there is an arrowhead at $i$ and $i\in\ant(C)$ in $\alpha_{CMG}(G;M,\varnothing)$.

By Lemma \ref{lem:1} and the fact that the anterior sets do not change at this step, the statement above is equivalent to  no arrowhead pointing to $i$ or  an arrowhead pointing to $i$ only when $i\in\ant(C)$ in the graph generated before applying step 2 of Algorithms \ref{alg:2} to $G$.

%
%
The result then follows from Lemma \ref{lem:vu} for the corresponding walk in $\alpha_{CMG}(\alpha_{CMG}(G;\varnothing,C);M,\varnothing)$.

\end{proof}
\begin{proof}[Proof of Proposition \ref{prop:2vnn}]
{\bf We first prove that every CG $G$ is mapped into $\mathcal{H}$:} By propositions \ref{prop:2}, \ref{prop:3}, and \ref{prop:l}, we conclude that the generated graphs are CMGs. By Proposition \ref{prop:2vn}, we know that $H=\alpha_{CMG}(G;M,\varnothing)$ is in $\mathcal{H}$. We need to prove that $H$ is mapped into $\mathcal{H}$ by conditioning.

Suppose that there is a collider trislide $\pi$ of form $k\arc i\ful\dots\ful j\fla l$ in the generated graph $\alpha_{CMG}(G;M,C)$. By Lemma \ref{lem:20}, the lines on $\pi$ exist in $H$. By Lemma \ref{lem:2n}, instead of the $lj$ arrow and the $ki$ arc, there are walks $\pi_1$ and $\pi_2$, respectively, as described in the lemma, in $H$. Consider the node $r$ adjacent to the endpoint section containing $j$ on $\pi_1$, and the node $h$ that is the other endpoint of the endpoint section containing $i$ on $\pi_2$. (Notice that $r$ may be $j$ and $h$ may be $i$.)

Since $H$ is in  $\mathcal{H}$, there is an arc (or an arrow if possibly $h=l$) between $r$ and $h$. Now the walk containing the subwalk of $\pi_1$ between $l$ and $r$, the $rh$ arc, and the subsection on $\pi_2$ between $h$ and $i$ satisfies the conditions of the walk described in Lemma \ref{lem:2n}. Hence, by this lemma, there is an arrow from $l$ to $i$ in  $\alpha_{CMG}(G;M,C)$.

%
%

If there is a collider trislide of form $k\arc i\ful\dots\ful j\arc l$ in the generated graph then by the same argument as that in the previous paragraph (and considering the fact that $k,l\notin S$), there are $il$ and $kj$ arcs in the generated graph. In addition, this time the walk containing the subwalk of $\pi_1$ between $j$ and $r$, the $rh$ arc, and the subsection on $\pi_2$ between $h$ and $i$ satisfies the conditions of the walk described in Lemma \ref{lem:2n}. Hence,  there is an arc between $j$ and $i$ in  $\alpha_{CMG}(G;M,C)$.

{\bf We now prove that the function is surjective:} by Proposition \ref{prop:2vn}, after marginalization, CGs are surjectively mapped onto $\mathcal{H}$. Thus, by letting $C=\varnothing$, Proposition \ref{prop:l}, and the fact that $\alpha_{CMG}(G;\varnothing,\varnothing)=G$, CGs are surjectively mapped onto $\mathcal{H}$ after marginalization and conditioning.
\end{proof}
\begin{proof}[Proof of Proposition \ref{prop:5}]
By Propositions \ref{prop:2} and \ref{prop:3}, we know that, after step 2 of Algorithm \ref{alg:4}, we obtain a CMG. Steps 3 and 4 do not generate a semi-directed cycle with an arrow by generating an arrow from $j$ to $i$: This is because if, for contradiction, that is the case then in the previous iteration of step 4, $j\in\ant(k)$ and $k\in\ant(i)$  which imply that $j\in\ant(i)$, and, in the previous iteration of step 3, $j\in\ant(i)$. This is a contradiction since it means by induction that the  semi-directed cycle with an arrow exists in the generated graph after applying step 2.

Step 5 obviously removes all arcs with one endpoint that is an anterior of the other endpoint. This step also does not generate semi-directed cycles with an arrow by replacing an arc $ij$ by an arrow from $j$ to $i$ or an $ij$ line: this is because if, for contradiction, that is the case then $j\in\ant(i)$ in the generated graph after applying step 4, which is a contradiction since it means by induction that the semi-directed cycle with an arrow exists in this graph.
\end{proof}
\begin{proof}[Proof of Lemma \ref{lem:30}]
We show that at every step of Algorithm \ref{alg:4}, a semi-directed path from $i$ to $j$ remain semi-directed and vice versa. For step 3 of the algorithm, the result is clear since the generating path of an arrow from $h$ to $l$ is semi-directed from $h$ to $l$. For step 4, this is correct as well since there is a node $k$ on the generating path such that $k\in\ant(l)$, and, on the generating path, $h\in\ant(k)$. This is also true for step 5 since if an arc turns into an arrow from $h$ to $l$ then $h$ is already an anterior of $l$.
\end{proof}
\begin{proof}[Proof of Lemma \ref{lem:31}]
{\bf First, we prove the first claim:}

{\bf ($\Rightarrow$)} Suppose that there is an $ij$ edge in $\alpha_{CMG.AnG}(H)$.
We see how this edge changes by steps of Algorithm \ref{alg:4}:

{\bf Before applying step 5:} There is still an edge between $i$ and $j$.

{\bf Before applying step 4:}
Instead of an arrow or an arc $ij$ at some iteration of this step of the algorithm, there may be a path between $i$ and $j$, consisting of one inner collider section and with inner nodes, say, in $\ant(i)$. By any other iteration, the arrow or the arc $kl$ might be replaced by another such path. By this replacement, we obtain a path (by discarding the intersection of lines) with all inner sections to be collider. Notice that by Lemma \ref{lem:30}, at no iteration the anterior set of the endpoints changes. In addition, regardless of whether inner nodes of the path between $k$ and $l$ are anteriors of $k$ or $l$, all inner nodes are anteriors of $i$. By an inductive argument, we finally obtain a subprimitive inducing path from $j$ to $i$.

{\bf In $H$:} By replacements of the arrow and arcs in step 3 of the algorithm, only sections become larger and inner nodes remain anteriors of an endpoint. If an endpoint of the arrow or arc is $i$ or $j$ then an endpoint section of the generated walk is not a single element and there is a node $h$ such that $h\in\ant(i)\cap \spo(i)$ or $h\in\ant(j)\cap \spo(j)$ respectively;  otherwise the endpoint sections are single elements. In the former case, we add $\langle i,h,i\rangle$ to the walk; and similarly for $j$.

{\bf ($\Leftarrow$)} Suppose that there is a subprimitive inducing walk $\pi$ from $j$ to $i$ in $H$. Consider the trislide $\rho$ containing $i$. First suppose that  the endpoints of $\rho$ are a single element $i$ (i.e.\ $\rho=\langle i,l,i\rangle$, where $l\in\ant(i)$). Consider the path $\langle k,\rho'\rangle$, where $i$ is an endpoint of the section $\rho'$ adjacent to $\rho$ and there is an arc between $k$ and the other endpoint of $\rho'$ (or possibly an arrow if $k=j$). By step 3 of Algorithm \ref{alg:4}, we can replace this path by an arc (or an arrow).

 By step 4 of the algorithm we obtain an arc instead of this trislide. By considering the trislide containing $i$ after the replacement, we have that inner nodes of the trislide are in $\ant(i)$. By repeating this argument we obtain an $ij$ edge.

{\bf We now prove the second claim:} If $j\in\ant(i)$ in $H$ then, by step 5 of the algorithm, there is no arrowhead at $j$ on the $ij$ edge in $\alpha_{CMG.AnG}(H)$.
If $j\not\in\ant(i)$ in $H$ then, by Lemma \ref{lem:30}, $j\not\in\ant(i)$ after applying step 4 of the algorithm. Hence, step 5 is not applicable. The result then follows from the fact that steps 3 and 4 generate endpoint-identical edges.
\end{proof}
\begin{proof}[Proof of Lemma \ref{lem:3}]
By Lemma \ref{lem:31}, it is enough  to prove that (1) there is a subprimitive inducing walk from $i$ to $j$ in $\alpha_{CMG}(\alpha_{CMG.AnG}(H);M,C)$ with single-element endpoint sections  if and only if there is an endpoint-identical walk of the same type from $i$ to $j$ in $\alpha_{CMG}(H;M,C)$; (2) $j\in \ant(i)$ in $\alpha_{CMG}(\alpha_{CMG.AnG}(H);M,C)$ if and only if $j\in \ant(i)$ in $\alpha_{CMG}(H;M,C)$.

{\bf Proving (1):} By Lemma \ref{lem:vu}, every edge on the subprimitive inducing walk $\pi$ from $i$ to $j$ in $\alpha_{CMG}(H;M,C)$ can be replaced by the described walk in the lemma. Denote the new walk by $\pi'$ in $H$. Notice that if a replaced subwalk is not endpoint-identical to the original edge then an endpoint $k$ of the edge should be in $\ant(C)$ in $H$, which means that $k$ is on a non-collider inner section on $\pi$ (or is an endpoint with no arrowheads pointing to it), but this is impossible. Therefore, all such edge-replacements are endpoint identical. In addition, by Lemma \ref{lem:an1}, if a node $h$ is in $\ant(j)$ in $\alpha_{CMG}(H;M,C)$ then $h\in\ant(C\cup\{j\})$ in $H$.

These imply that there is a subprimitive inducing walk from $i$ to $j$ with the mentioned properties in $\alpha_{CMG}(H;M,C)$ if and only if in $H$ there is a walk between $i$ and $j$ on which (i) all nodes on collider sections are in $C\cup\ant(C)\cup\{j\}$; (ii) (a) all nodes on non-collider sections are in $M$, or (b) on non-collider sections, one endpoint is in $M$ and also either a child of a node in $M$ or a spouse of a node in $C\cup\ant(C)$, and the other endpoint has an arrowhead at it from the adjacent node on the walk. In addition, the two walks  are endpoint-identical except when there is an arrowhead at the endpoint section containing $i$ (or $j$), and $i\in \ant(C)$ (or $j\in \ant(C)$) in $H$.

%
%
%

Now by using Lemma \ref{lem:30}, we have that $i\in\ant(C)$ in $H$ if and only if $i\in\ant(C)$ in $\alpha_{CMG.AnG}(H)$. Therefore, since the same statements as above hold also for  $\alpha_{CMG}(\alpha_{CMG.AnG}(H);M,C)$ and $\alpha_{CMG.AnG}(H)$, and  in order to complete the proof, we need to show that there is a walk between $i$ and $j$ in $H$ with the two mentioned properties if and only if there is an endpoint-identical walk $\pi_0$ of the same type  between $i$ and $j$  in  $\alpha_{CMG.AnG}(H)$:

To prove this, it is enough to  show that by placing the walks  described in Lemma \ref{lem:31} in place of the edges of $\pi_0$, the form of $\pi_0$ does not change: Without loss of generality, suppose that $\pi_0$ is a shortest walk of the described form, and an $rs$ edge on $\pi_0$ has been replaced by a subprimitive inducing walk $\varpi$ from $r$ to $s$. The newly added sections are all collider. Because of transitivity of anteriors, and since the inner nodes of $\varpi$ are anteriors of $s$, they stay is $\ant(C\cup\{j\})$. It is now enough  to only check the sections containing $r$ and $s$ on $\pi_0$. Firstly, it is easy to see by Lemma \ref{lem:31} that the type of these sections do not change regardless of whether they are single elements on $\varpi$.

Secondly, if the $rs$ edge and $\varpi$ are endpoint-identical then theses sections remain of the same type. This completes the proof by using Lemma \ref{lem:30}.

If these are not endpoint-identical then $s\in\ant(r)$. A problem only may arise when the section containing $s$ is a non-collider in $\alpha_{CMG.AnG}(H)$ but a collider in $H$. If, for contradiction, this is the case then there is an arrow to $s$ from the other adjacent node $q$ to $s$ on $\pi_0$.  In addition, since all inner nodes of $\varpi$ are anteriors of $s$, they are anteriors of $r$, and hence in $H$, $\langle \varpi,q\rangle$ is a subprimitive inducing walk from $q$ to $r$, and hence $\pi_0$ is not a shortest walk, a contradiction. This completes the proof of this section.

{\bf Proving (2):} Consider a semi-directed walk $\pi$ in $\alpha_{CMG}(\alpha_{CMG.AnG}(H);M,C)$ from $j$ to $i$. Since every edge is a subprimitive inducing walk, lines on $\pi$ remain the same, and instead of an arrow from $k$ to $l$ on $\pi$ we may have a  subprimitive inducing walk from $k$ to $l$. It is easy to observe that $k\in\ant(l)$, and by an inductive argument, we obtain the result.

The proof of other direction uses exactly the same argument (although, in fact, edges remain edges in this case).
\end{proof}
\begin{proof}[Proof of Proposition \ref{prop:5vn}]
{\bf First we prove that every CG $G$ is mapped into $\mathcal{K}$:} By Proposition \ref{prop:5}, we know that $\alpha_{AnG}$ maps CGs into $\mathcal{ANG}$. By Proposition \ref{prop:2vnn}, we know that after applying steps 1 and 2 of Algorithm \ref{alg:4}, a CG $G$ is mapped into $\mathcal{H}$, defined in Proposition \ref{prop:2vn}. We need to prove that after applying steps 3, 4, and 5 of Algorithm \ref{alg:4}, a CMG $H\in \mathcal{H}$ is mapped into $\mathcal{K}$.

{\bf Suppose that there is a trislide $\pi=k\arc i\ful\dots\ful j\fla l$ in the generated graph}:
%
%
By Lemma \ref{lem:31}, there is a subprimitive inducing walk from $l$ to $j$ in $H$. Denote the node on this walk adjacent to $j$ by $q$. The $jq$ edge is an arc unless $l=q$, in which case it is an arrow from $q$ to $j$. Since lines are not generated by Algorithm \ref{alg:4}, and since $H\in \mathcal{H}$, there is an $iq$ arc or an arrow from $l$ to $i$.

In the generated graph, $j\in\ant(i)$, and there is a subprimitive inducing walk from $l$ to $i$ that goes through the subprimitive inducing walk from $l$ to $j$, the section from $j$ to $i$, the $iq$ edge, the $jq$ edge, and again the section between $j$ and $i$. Hence, again by Lemma \ref{lem:31}, there is an edge between $l$ and $i$. This edge can only be an arrow from $l$ to $i$ since otherwise there is a semi-directed cycle or an arc with one endpoint that is an anterior of the other endpoint in the generated anterial graph.

%
%
%
 {\bf Suppose that there is a trislide $\pi=k\arc i\ful\dots\ful j\arc l$ in the generated graph:} It holds that $l\notin\ant(i)$ since otherwise $l\in\ant(j)$, which is impossible due to the existence of an arrowhead at $l$. This fact together with the same argument as that in the previous paragraphs implies that there is an $il$ arc in the generated graph. By the symmetry on the trislide we also conclude that  there is a $jk$ arc in the generated graph. In addition, by what we proved in the previous paragraphs, there is a tripath $q'\arc i\ful\dots\ful j\arc q$ in $H$, which implies that
there is an $ij$ arc in $H$. This arc turns into a line by step 5 since $i$ and $j$ are anteriors of one another.

{\bf We now prove that the function is surjective:} Consider an arbitrary graph $K\in\mathcal{K}$. We prove that there exists an $H\in\mathcal{H}$ such that $\alpha_{CMG.AnG}(H)=K$, i.e.\ by applying steps 3, 4, and 5 of Algorithm \ref{alg:4} to $H$, we obtain $K$. This completes the proof since $\alpha_{CMG}$ is surjective onto $\mathcal{H}$, and $\alpha_{AnG}=\alpha_{CMG.AnG}\circ\alpha_{CMG}$.

If $K$ does not contain a trislide of form $\pi=k\arc i\ful\dots\ful j\arc l$ then $K\in\mathcal{H}$, and we simply let $H=K$. Since $\alpha_{AnG}$ does not change anterial graphs, we are done.

If $K$ does contain a trislide $\pi$ of the mentioned form then there is the $ij$ line in $K$. Now let $H$ be $K$, but with an arc between $i$ and $j$ instead of the existing line. We have that $H\in\mathcal{H}$. Denote also the section between $i$ and $j$ by $\rho$.

By Lemma \ref{lem:31}, the $ij$ arc turns into a line and clearly no other edge changes its type in $\alpha_{CMG.AnG}(H)$. Hence, it is enough to show that no other edge is generated. If the $ij$ arc is part of any subprimitive inducing walk except when $i$ or $j$ is an endpoint then it can be replaced by $\rho$ to obtain another primitive inducing walk. If $i$ or $j$ is an endpoint then, by how $H$ is constructed, the possible arrows or lines that can be generated already exist in $H$. This completes the proof.
%
%
\end{proof}
\begin{proof}[Proof of Theorem \ref{thm:5}]
By Theorem \ref{thm:4}, it is enough to prove that $A\dse_cB\cd C_1$ in $\alpha_{AnG}(G;M,C)$ if and only if $A\dse_cB\cd C_1$ in $\alpha_{CMG}(G;M,C)$.

Since Steps 1 and 2 of Algorithm \ref{alg:4} generate $\alpha_{CMG}(G;M,C)$, we need to prove that there is a $c$-connecting walk in a chain mixed graph $H$ if and only if there is a $c$-connecting walk after applying steps 3, 4, and 5 of the algorithm to $H$.

{\bf ($\Rightarrow$)} Suppose that there is a $c$-connecting walk $\pi$ given $C_1$ between $i$ and $j$ in $H$. After applying steps 3 and 4, $\pi$ is intact. If an arc $kl$ is replaced by an arrow from $k$ to $l$ or a $kl$ line, in step 5 of the algorithm then we have the two following cases:

{\bf 1)} If $k$ is on a non-collider section on $\pi$ by using the $kl$ arrow or line instead of arc, one obtains a $c$-connecting walk.

{\bf 2)} Suppose that $k$ is an endpoint of a collider section $\rho$ and there is $\pi_1=\langle h,\rho,l\rangle$ on $\pi$. By Lemma \ref{lem:0}, one can assume that $\rho$ is a path. By Lemma \ref{lem:30}, $k\in\ant(l)$. If $h\neq l$ then by step 4, there is an endpoint-identical $hl$ edge to $\pi_1$. One can now use the $hl$ edge instead of $\pi_1$ to obtain a $c$-connecting walk. If $h=l$ then $\rho$ can be considered to be the single node $k$. Now if $h$ is on a non-collider section then we can easily skip $k$ to obtain a $c$-connecting path. If $h$ is an endpoint of a collider section $\rho'$ then from $\pi_2=\langle q,\rho',k\rangle$ and by using step 3 of the algorithm, we obtain an endpoint-identical $qh$ edge, which can be replaced by $\pi_2$ to obtain a $c$-connecting path. This, by an inductive argument, implies the result.

{\bf ($\Leftarrow$)} Suppose that there is a $c$-connecting walk $\pi$ given $C_1$ between $i$ and $j$ in $\alpha_{CMG.AnG}(H)$, which is graph $H$ after applying steps 3, 4, and 5 of Algorithm \ref{alg:4}.

For every edge on $\pi$, by Lemma \ref{lem:31}, there exists a subprimitive inducing walk in $H$ between the same endpoints. We replace all the edges on $\pi$ by these walks and call the generated walk $\pi'$. Notice that it can be shown that regardless of the choice of $C$, a subprimitive inducing walk is $c$-connecting itself. Hence, if the replaced subwalk of $\pi'$ by an edge is endpoint-identical to the original edge then it does not affect the $c$-connectivity of $\pi'$. We, therefore, need to check the case where the generated walk is not endpoint-identical to the edge.

Suppose that this is the case for the edge $ij$ in $\alpha_{CMG.AnG}(H)$ replaced by a subprimitive inducing walk $\varpi$ from $j$ to $i$. By the lemma, we have that either $j\in\ant(i)$ or $i\in\ant(j)$ in $H$, in which cases there is no arrowhead at $j$ or $i$ on the $ij$ edge respectively.

Assume that $j\in\ant(i)$. We need to consider the case where $ij$ is an arrow from $j$ to $i$, and $j$ is not in $C$, but there is an arrowhead at $j$ on $\varpi$. Denote the semi-directed walk from $j$ to $i$ by $\tau$. If no node on $\tau$ is in $C$ then we replace $\varpi$ by $\tau$ to obtain a $c$-connecting walk. Otherwise, consider the closest node $k\in C$ on $\tau$ to $j$. The walk consisting of the subwalk of $\tau$ from $j$ to $k$, the same subwalk in the reverse direction (from $k$ to $j$), and $\varpi$ is now $c$-connecting since $j$ is on non-collider sections, except when $j$ and $k$ are on the same subsection of $\tau$ (which is still fine).

The case where $i\in\ant(j)$ follows the exact same argument.
\end{proof}
\bibliographystyle{imsart-nameyear}
\bibliography{bib}

\begin{thebibliography}{28}

\bibitem[\protect\citeauthoryear{Andersson, Madigan and Perlman.}{2001}]{and01}
\begin{barticle}[author]
\bauthor{\bsnm{Andersson},~\bfnm{Steen~A.}\binits{S.~A.}},
  \bauthor{\bsnm{Madigan},~\bfnm{David}\binits{D.}} \AND
  \bauthor{\bsnm{Perlman.},~\bfnm{Michael~D.}\binits{M.~D.}}
(\byear{2001}).
\btitle{Alternative Markov Properties for Chain Graphs}.
\bjournal{Scand. J. Stat.}
\bvolume{28}
\bpages{33-85}.
\end{barticle}
\endbibitem

\bibitem[\protect\citeauthoryear{Cox and Wermuth}{1993}]{cox93}
\begin{barticle}[author]
\bauthor{\bsnm{Cox},~\bfnm{D.~R.}\binits{D.~R.}} \AND
  \bauthor{\bsnm{Wermuth},~\bfnm{N.}\binits{N.}}
(\byear{1993}).
\btitle{Linear dependencies represented by chain graphs (with discussion)}.
\bjournal{Stat. Sci.}
\bvolume{8}
\bpages{204--218; 247--277}.
\end{barticle}
\endbibitem

\bibitem[\protect\citeauthoryear{Drton}{2009}]{drt09}
\begin{barticle}[author]
\bauthor{\bsnm{Drton},~\bfnm{M.}\binits{M.}}
(\byear{2009}).
\btitle{Discrete chain graph models}.
\bjournal{Bernoulli}
\bvolume{15}
\bpages{736--753}.
\end{barticle}
\endbibitem

\bibitem[\protect\citeauthoryear{Evans and Richardson}{2014}]{eva14}
\begin{barticle}[author]
\bauthor{\bsnm{Evans},~\bfnm{Robin~J.}\binits{R.~J.}} \AND
  \bauthor{\bsnm{Richardson},~\bfnm{Thomas~S.}\binits{T.~S.}}
(\byear{2014}).
\btitle{Markovian acyclic directed mixed graphs for discrete data}.
\bjournal{Ann. Statist.}
\bvolume{42}
\bpages{1452-1482}.
\end{barticle}
\endbibitem

\bibitem[\protect\citeauthoryear{Frydenberg}{1990}]{fry90}
\begin{barticle}[author]
\bauthor{\bsnm{Frydenberg},~\bfnm{M.}\binits{M.}}
(\byear{1990}).
\btitle{The chain graph {M}arkov property}.
\bjournal{Scand. J. Stat.}
\bvolume{17}
\bpages{333--353}.
\end{barticle}
\endbibitem

\bibitem[\protect\citeauthoryear{Geiger et~al.}{2001}]{gei01}
\begin{barticle}[author]
\bauthor{\bsnm{Geiger},~\bfnm{D.}\binits{D.}},
  \bauthor{\bsnm{Heckerman},~\bfnm{D.}\binits{D.}},
  \bauthor{\bsnm{King},~\bfnm{H.}\binits{H.}} \AND
  \bauthor{\bsnm{Meek},~\bfnm{C.}\binits{C.}}
(\byear{2001}).
\btitle{Stratified exponential families: Graphical models and model selection}.
\bjournal{Ann. Statist.}
\bvolume{29}
\bpages{505-529}.
\end{barticle}
\endbibitem

\bibitem[\protect\citeauthoryear{Kiiveri, Speed and Carlin}{1984}]{kii84}
\begin{barticle}[author]
\bauthor{\bsnm{Kiiveri},~\bfnm{H.}\binits{H.}},
  \bauthor{\bsnm{Speed},~\bfnm{T.~P.}\binits{T.~P.}} \AND
  \bauthor{\bsnm{Carlin},~\bfnm{J.~B.}\binits{J.~B.}}
(\byear{1984}).
\btitle{Recursive causal models}.
\bjournal{J. Aust. Math. Soc., Ser. A}
\bvolume{36}
\bpages{30--52}.
\end{barticle}
\endbibitem

\bibitem[\protect\citeauthoryear{Koster}{2002}]{kos02}
\begin{barticle}[author]
\bauthor{\bsnm{Koster},~\bfnm{J.~T.~A.}\binits{J.~T.~A.}}
(\byear{2002}).
\btitle{Marginalizing and conditioning in graphical models}.
\bjournal{Bernoulli}
\bvolume{8}
\bpages{817--840}.
\end{barticle}
\endbibitem

\bibitem[\protect\citeauthoryear{Lauritzen}{1996}]{lau96}
\begin{bbook}[author]
\bauthor{\bsnm{Lauritzen},~\bfnm{S.~L.}\binits{S.~L.}}
(\byear{1996}).
\btitle{Graphical Models}.
\bpublisher{Clarendon Press}, \baddress{Oxford, United Kingdom}.
\end{bbook}
\endbibitem

\bibitem[\protect\citeauthoryear{Lauritzen and Spiegelhalter}{1988}]{lau88}
\begin{barticle}[author]
\bauthor{\bsnm{Lauritzen},~\bfnm{S.~L.}\binits{S.~L.}} \AND
  \bauthor{\bsnm{Spiegelhalter},~\bfnm{D.~J.}\binits{D.~J.}}
(\byear{1988}).
\btitle{Local computations with probabilities on graphical structures and their
  application to expert systems}.
\bjournal{J. Roy. Statis. Society B}
\bvolume{50}
\bpages{157-224}.
\end{barticle}
\endbibitem

\bibitem[\protect\citeauthoryear{Lauritzen and Wermuth}{1989}]{lau89}
\begin{barticle}[author]
\bauthor{\bsnm{Lauritzen},~\bfnm{S.~L.}\binits{S.~L.}} \AND
  \bauthor{\bsnm{Wermuth},~\bfnm{N.}\binits{N.}}
(\byear{1989}).
\btitle{Graphical models for association between variables, some of which are
  qualitative and some quantitative}.
\bjournal{Ann. Statist.}
\bvolume{17}
\bpages{31--57}.
\end{barticle}
\endbibitem

\bibitem[\protect\citeauthoryear{Marchetti and Lupparelli}{2011}]{mar11}
\begin{barticle}[author]
\bauthor{\bsnm{Marchetti},~\bfnm{Giovanni~M.}\binits{G.~M.}} \AND
  \bauthor{\bsnm{Lupparelli},~\bfnm{Monia}\binits{M.}}
(\byear{2011}).
\btitle{Chain graph models of multivariate regression type for categorical
  data}.
\bjournal{Bernoulli}
\bvolume{17}
\bpages{827-844}.
\end{barticle}
\endbibitem

\bibitem[\protect\citeauthoryear{Pe\~{n}a}{2009}]{pen09}
\begin{barticle}[author]
\bauthor{\bsnm{Pe\~{n}a},~\bfnm{Jose~M.}\binits{J.~M.}}
(\byear{2009}).
\btitle{Faithfulness in chain graphs: The discrete case}.
\bjournal{Int. J. Approx. Reason.}
\bvolume{50}
\bpages{1306 - 1313}.
\end{barticle}
\endbibitem

\bibitem[\protect\citeauthoryear{Pe\~{n}a}{2011}]{pen11}
\begin{binproceedings}[author]
\bauthor{\bsnm{Pe\~{n}a},~\bfnm{Jose~M.}\binits{J.~M.}}
(\byear{2011}).
\btitle{Faithfulness in Chain Graphs: The Gaussian Case.}
In \bbooktitle{Proceedings of the 14th International Conference on Artificial
  Intelligence and Statistics (AISTATS 2011)}
\bvolume{15}
\bpages{588-599}.
\bpublisher{JMLR.org}.
\end{binproceedings}
\endbibitem

\bibitem[\protect\citeauthoryear{Pe\~{n}a}{2014}]{pen14}
\begin{barticle}[author]
\bauthor{\bsnm{Pe\~{n}a},~\bfnm{Jose~M.}\binits{J.~M.}}
(\byear{2014}).
\btitle{Marginal {AMP} chain graphs}.
\bjournal{Int. J. Approx. Reason.}
\bvolume{55}
\bpages{1185-1206}.
\end{barticle}
\endbibitem

\bibitem[\protect\citeauthoryear{Pearl}{2009}]{pea09}
\begin{bbook}[author]
\bauthor{\bsnm{Pearl},~\bfnm{Judea}\binits{J.}}
(\byear{2009}).
\btitle{Causality: Models, Reasoning and Inference},
\bedition{2nd} ed.
\bpublisher{Cambridge University Press}, \baddress{New York, NY, USA}.
\end{bbook}
\endbibitem

\bibitem[\protect\citeauthoryear{Richardson}{2003}]{ric03}
\begin{barticle}[author]
\bauthor{\bsnm{Richardson},~\bfnm{T}\binits{T.}}
(\byear{2003}).
\btitle{Markov Properties for Acyclic Directed Mixed Graphs}.
\bjournal{Scand. J. Stat.}
\bvolume{30}
\bpages{145--157}.
\bdoi{10.1111/1467-9469.00323}
\end{barticle}
\endbibitem

\bibitem[\protect\citeauthoryear{Richardson and Spirtes}{2002}]{ric02}
\begin{barticle}[author]
\bauthor{\bsnm{Richardson},~\bfnm{T.~S.}\binits{T.~S.}} \AND
  \bauthor{\bsnm{Spirtes},~\bfnm{P.}\binits{P.}}
(\byear{2002}).
\btitle{Ancestral graph {M}arkov models}.
\bjournal{Ann. Statist.}
\bvolume{30}
\bpages{962--1030}.
\end{barticle}
\endbibitem

\bibitem[\protect\citeauthoryear{Sadeghi}{2013}]{sad13}
\begin{barticle}[author]
\bauthor{\bsnm{Sadeghi},~\bfnm{Kayvan}\binits{K.}}
(\byear{2013}).
\btitle{Stable mixed graphs}.
\bjournal{Bernoulli}
\bvolume{19}
\bpages{2330-2358}.
\end{barticle}
\endbibitem

\bibitem[\protect\citeauthoryear{Sadeghi}{2015}]{sad15s}
\begin{barticle}[author]
\bauthor{\bsnm{Sadeghi},~\bfnm{Kayvan}\binits{K.}}
(\byear{2015}).
\btitle{Supplement to {``}Marginalization and conditioning for {LWF} chain
  graphs{"}}.
\end{barticle}
\endbibitem

\bibitem[\protect\citeauthoryear{Shpitser and Pearl}{2008}]{shp08}
\begin{binproceedings}[author]
\bauthor{\bsnm{Shpitser},~\bfnm{I.}\binits{I.}} \AND
  \bauthor{\bsnm{Pearl},~\bfnm{J.}\binits{J.}}
(\byear{2008}).
\btitle{Dormant independence}.
In \bbooktitle{Proceedings of the twenty-third AAAI Conference on Artificial
  Inteligence}
\bpages{1081-1087}.
\bpublisher{AAAI Press}.
\end{binproceedings}
\endbibitem

\bibitem[\protect\citeauthoryear{Studeny}{1998}]{stu98}
\begin{binproceedings}[author]
\bauthor{\bsnm{Studeny},~\bfnm{Milan}\binits{M.}}
(\byear{1998}).
\btitle{Bayesian Networks from the Point of View of Chain Graphs}.
In \bbooktitle{UAI}
\bpages{496-503}.
\bpublisher{Morgan Kaufmann}, \baddress{San Francisco, CA}.
\end{binproceedings}
\endbibitem

\bibitem[\protect\citeauthoryear{Studeny}{2005}]{stu05}
\begin{bbook}[author]
\bauthor{\bsnm{Studeny},~\bfnm{M.}\binits{M.}}
(\byear{2005}).
\btitle{Probabilistic Conditional Independence Structures}.
\bpublisher{Springer-Verlag}, \baddress{London, United Kingdom}.
\end{bbook}
\endbibitem

\bibitem[\protect\citeauthoryear{Studeny and Bouckaert}{1998}]{stub98}
\begin{barticle}[author]
\bauthor{\bsnm{Studeny},~\bfnm{M.}\binits{M.}} \AND
  \bauthor{\bsnm{Bouckaert},~\bfnm{R.~R.}\binits{R.~R.}}
(\byear{1998}).
\btitle{On chain graph models for description of conditional independence
  structures}.
\bjournal{Ann. Statist.}
\bvolume{26}
\bpages{1434--1495}.
\end{barticle}
\endbibitem

\bibitem[\protect\citeauthoryear{Verma and Pearl}{1990}]{ver90}
\begin{binproceedings}[author]
\bauthor{\bsnm{Verma},~\bfnm{T.}\binits{T.}} \AND
  \bauthor{\bsnm{Pearl},~\bfnm{J.}\binits{J.}}
(\byear{1990}).
\btitle{Equivalence and synthesis of causal models}.
In \bbooktitle{Proceedings of the Sixth Conference on Uncertainty in Artificial
  Intelligence (UAI-90)}
\bpages{220--227}.
\end{binproceedings}
\endbibitem

\bibitem[\protect\citeauthoryear{Wermuth}{2011}]{wer11}
\begin{barticle}[author]
\bauthor{\bsnm{Wermuth},~\bfnm{N.}\binits{N.}}
(\byear{2011}).
\btitle{Probability distributions with summary graph structure}.
\bjournal{Bernoulli}
\bvolume{17}
\bpages{845--879}.
\end{barticle}
\endbibitem

\bibitem[\protect\citeauthoryear{Wermuth and Sadeghi}{2012}]{wers11}
\begin{barticle}[author]
\bauthor{\bsnm{Wermuth},~\bfnm{Nanny}\binits{N.}} \AND
  \bauthor{\bsnm{Sadeghi},~\bfnm{Kayvan}\binits{K.}}
(\byear{2012}).
\btitle{Sequences of regressions and their independences}.
\bjournal{TEST}
\bvolume{21}
\bpages{215-252 and 274-279}.
\end{barticle}
\endbibitem

\bibitem[\protect\citeauthoryear{Wermuth, Wiedenbeck and Cox}{2006}]{wer06}
\begin{barticle}[author]
\bauthor{\bsnm{Wermuth},~\bfnm{N.}\binits{N.}},
  \bauthor{\bsnm{Wiedenbeck},~\bfnm{M.}\binits{M.}} \AND
  \bauthor{\bsnm{Cox},~\bfnm{D.~R.}\binits{D.~R.}}
(\byear{2006}).
\btitle{Partial inversion for linear systems and partial closure of
  independence graphs}.
\bjournal{BIT}
\bvolume{46}
\bpages{883--901}.
\end{barticle}
\endbibitem

\end{thebibliography}

\end{document}